\documentclass[preprint,preprintnumbers,amsmath,amssymb]{revtex4}

\usepackage{graphicx}
\usepackage{dcolumn}
\usepackage{bm}

\begin{document}

\title{Chaotic oscillations in a map-based model of neural activity \\}

\author{M. Courbage$^{a}$}
\author{ V.I. Nekorkin$^{b}$}
\author{L.V. Vdovin$^{b}$}
\affiliation{ (a) Laboratoire Mati\`{e}re et syst\`{e}mes Complexes(MSC)\\ UMR 7057 CNRS et
Universit\'e Paris 7-Denis Diderot\\ Batiment Condorcet 75205 Paris Cedex 13, France\\ (b)
Institute of Applied Physics of RAS,\\ Nizhny Novgorod State University, Nizhny Novgorod,
Russia\\}

\date{\today}

\begin{abstract}
We propose a discrete time dynamical system (a map) as phenomenological
model of excitable and spiking-bursting neurons. The model is a
discontinuous two-dimensional map. We find condition under which
this map has an invariant region on the phase plane, containing
chaotic attractor. This attractor creates chaotic spiking-bursting
oscillations of the model. We also  show various regimes of  other
neural activities (subthreshold oscillations, phasic spiking etc.)
derived from the proposed model.
\end{abstract}

\pacs{ XXX }

\maketitle

{ \bf The observed types of neural activity are extremely various.
 A single neuron may display different regimes of activity
 under different
neuromodulatory conditions. A neuron is said to produce excitable
mode if a "superthreshold" synaptic input  evokes a post-synaptic
potential in form of single spikes, which is an order of magnitude
larger than the input amplitude. While a "subthreshold" synaptic input
evokes   post-synaptic potentials of the same order.
Under some conditions a single spike can be generated with
arbitrary low frequency, depending on the strength of the applied
current.  It is called spiking regime. An important regime of
neural activity is bursting oscillations where clusters of spikes  occur periodically or chaotically,
separated by phases of quiescence. Other important observed regimes are phasic spikes and 
bursts, subtreshold oscillations and tonic spiking.
Understanding dynamical mechanisms of such activity in biological
neurons has stimulated the development of models on several
levels of complexity.  To explain biophysical membrane processes
in a single cell, it is generally used ionic channel-based models
\cite{b:Kandel}. The prototype of those models is the
Hodgkin-Huxley system which was originally introduced in the
description of the membrane potential dynamics in the giant squid
axon. This is a high dimensional system of
nonlinear partial differential equations. Another class of neuron models
are the phenomenological models  which  mimic qualitatively 
some distinctive features of neural activity with few
differential equations. For example, the leaky integrate-and-fire model,
Hindmarsh-Rose and FitzHugh-Nagumo model etc. A new important
subclass of phenomenological models is the map-based systems.
Basically such models are designed with the aim of simulating
collective dynamics of large neuronal networks. The map-based
models possess at least the same features of Ordinary differential equations (ODE) models, and
have more simple intrinsic structure offering an advantage in
describing more complex dynamics. In order to model basic
regimes of neural activity we design new family of maps that are
two-dimensional and based on discrete  FitzHugh-Nagumo system in
which we introduce Heaviside step function. The discontinuity line
determines the excitation threshold of chaotic spiking-bursting
oscillations. For some domain of the parameters, we found on phase plane an invariant bounded
region containing chaotic attractor with spiking-bursting
activity. The interesting fact is that the dynamical mechanism,
leading to chaotic behavior of our two-dimensional map is induced
by one-dimensional Lorenz-like map. We demonstrate also that our
model can display rich gallery of regimes of neural activity such
as chaotic spiking, subthreshold oscillations, tonic spiking
etc.All these modes play important role in the information
processing in neural systems. }

\section{\label{sec:intro}Introduction \protect\\}
The nervous system is an extremely complex system \cite{b:Kandel}
comprising nerve  cells (or neurons) and gial cells. By
electrical and chemical synapses of different polarity neurons
form a great variety large-scale networks.
Therefore, modeling of brain's key functional properties is 
associated with study of collective activity of complex
neurobiological networks. Dynamical modeling approach
\cite{b:Rabinovich} is effective tool for the analysis of this
kind of networks. First of all this approach takes building
dynamical models of single neurons. On the one hand, such models
should describe large quantity of various dynamical modes of
neural activity (excitable, oscillatory, spiking, bursting, etc.).
This complexity is associated with the large number of
voltage-gated ion channels of neurons. It takes employment of
complex nonlinear dynamical systems given by  differential
equations. The canonical representative of this type of
models is Hodgkin-Huxley system. It describes dynamics of the
transport through membrane of neuron in detail. On the other hand,
to model neural network consisting of the large number of
interconnected units it is necessary to create  simplified
models for single neuron to avoid problems that are induced by
high dimension and nonlinearity. For example, one
which is commonly used in simulations is integrate-and-fire model
\cite{b:IandFModel}. It represents one-dimensional nonlinear
equation with some threshold rule. That is, if the variable of the model
crosses a critical value, then it is reset to new value and the
neuron is said to have fired. To solve the contradiction between
the requirements of  complexity and simplicity of neuron models
phenomenological models were introduced. They describe basic
properties of neuron dynamics, but these models do not take into
account the large number of voltage-gated ion channels of
neurons. As a rule they involve generalized variable which mimic
the dynamic of  some number of ionic currents at the same time.
The examples of this type models are FitzHugh-Nagumo
\cite{b:FitzHugh}, Hindmarsh-Rose \cite{b:HingmarshRose},
Morris-Lecar \cite{b:MorrisLekar}. They have the form
of  differential equations systems. However, there is another class of
phenomenological models of the neural activity. These are 
discrete-time models in form of point maps.  In the last decade
this kind of neural models has attracted much
attention\cite{b:Chialvo,b:Kinouchi, b:Kuva, b:DeVries}. For
example using a map-based approach Rulkov et. al.
\cite{b:RulkovTim} have studied dynamics of one- and -two
dimensional large-scale cortical networks. It has been found that
such map-based models produce spatiotemoral regimes similar to
those exhibited by Hodgkin-Huxley -like models.
\par
Neuron oscillatory activity can take a variety of forms
\cite{b:Traub}. One of the most interesting oscillatory regimes is
spiking-bursting oscillations regime, which is commonly observed in a wide
variety of neurons such as hippocampal pyramidal neurons, thalamic
neurons, pyloric dilator neurons etc. A burst is a series of three
or more action potential usually riding on a depolarizing wave. It
is believed that the bursting oscillations play crucial role in
informational transmitting and processing in neurons, facilitate
secretion of hormones and drive a muscle contraction. This
oscillation can be regular or chaotic depending on the
concentration of neuromodulators, currents and other control
parameters. Another interesting oscillatory regime is an 
oscillation of membrane potential below the excitation threshold,
so-called subthreshold oscillation. For example, these oscillations
with close to 10 Hz frequency are observed in olivo-cerebellar
system providing highly coordinated signals concerned with the
temporal organization of movement execution \cite{b:Llinas,
b:LlinasVortex} (see more discussion in the conclusion).
\par
The best known  spiking-bursting activity model is the Hindmarsh-Rose
system \cite{b:HingmarshRose}. It is three-dimensional ODE-based
system involving two nonlinear functions. Spiking-bursting dynamics of
map-based models has recently been investigated by Cazelles et.al
\cite{b:CazellisCourbage}, Rulkov \cite{b:RulkovSynchro,
b:RulkovBase}, Shilnikov and Rulkov
\cite{b:RulkovChaos1,b:RulkovChaos2}, Tanaka \cite{b:TanakaH}. A
piecewise linear two-dimensional map with a fast-slow dynamics was
introduced in \cite{b:CazellisCourbage}. It was shown that
depending on the connection (diffusively or reciprocally
synoptically), the model demonstrates several modes of cooperative
dynamics, among them phase synchronization. Two dimensional map is
used for modeling of spiking-bursting neural behavior of neuron
\cite{b:RulkovBase,b:RulkovChaos1,b:RulkovChaos2}. This map
contains one fast and one slow variable. The map is piecewise
nonlinear and has two lines of discontinuity on the phase plane.
Modification of this model is presented in \cite{b:RulkovChaos1}.
The further advancement of Rulkov model  is presented in
\cite{b:RulkovChaos2}. A quadratic function has been introduced
in the model. Using these modifications authors obtained the dynamical
regimes of subthreshold oscillation, corresponding to the
periodical oscillation of neuron's transmembrane potential below
the excitation threshold. In \cite{b:CourbageNekorkin} the
dynamics of two coupled piece-wise linear one-dimensional
monostable maps is investigated. The single map is associated with
Poincar\'e section of the FitzHugh-Nagumo neuron model. It is
found that a diffusive coupling leads to the appearance of chaotic
attractor. The attractor exists in an invariant region of phase
space bounded by the manifolds of the saddle fixed point and the
saddle periodic point. The oscillations from the chaotic attractor
have a spike-burst shape with anti-phase synchronized  spiking.
A map-based neuron model involving quasi-periodic oscillation
for generating the bursting activity has been suggested in
\cite{b:TanakaH}. Izhikevich and Hoppenstead have classified
\cite{b:Izhikevich} map-based one- and two-dimensional models of
bursting activity using bifurcation theory.
\par
Our goal here is to introduce  a new map-based model for
replication of many basic modes of neuron activity. The greater
part of our paper deals with regimes that mimic chaotic
spiking-bursting activity of one real biological neuron. We
construct a discontinuous two-dimensional map based on well-known
one-dimensional Lorenz-type map \cite{b:Afraimovich} and a discrete
version of the FitzHugh-Nagumo model\cite{b:FitzHugh}. This is the
system of two ODE:
\begin{eqnarray} \label{eq:ModelFHN}
     \dot{x} = F(x) - y \\
     \dot{y} = \varepsilon (x-J)
\end{eqnarray}
where $x$ is the membrane potential of the neuron and $y$ is the
recovery variable describing ionic currents, $F$ is a cubic
function of $x$
and $J$ is constant stimulus. This model takes into account the
excitability and regular oscillations of neuron, but not
spiking-bursting behavior. We shall introduce a discontinuity in
the discrete version for this purpose. We find conditions under
which this two-dimensional map has an invariant region on the
phase plane, containing chaotic attractor. In addition we show
that depending on the values of the parameters, our model can
produce phasic spiking and subthreshold oscillations also.
\par
The paper is organized as follows. In Sec. \ref{sec:model} we
describe the map-based model. Then in Sec. \ref{sec:OneDDyn} we
study one-dimensional dynamics in the case when the recovery
variable $y$ is fixed. In Sec. \ref{sec:TwoDDyn} we analyze the
relaxation two-dimensional dynamics of the model. Then in Sec.
\ref{sec:InvR} we find an invariant region bounding the chaotic
attractor in the phase plane of the model. In Sec.
\ref{sec:OtherModes} we observe other modes of neural activity
which could be simulated by using this model.

\section{\label{sec:model}A model for bursting neural cell \protect\\}
Let $f: R^2 \rightarrow R^2$ be a map $(x,y) \rightarrow
(\overline{x},\overline{y}) $  of the form
\begin{equation} \label{eq:stpMp}
     \begin{cases}
         \overline{x} = x + F(x) - y - \beta H(x-d) \\
         \overline{y}= y + \varepsilon(x-J)
     \end{cases},
\end{equation}
where the $x$-variable describes the evolution of the membrane
potential of the neuron, the $y$ - variable describes the dynamics of
the outward ionic currents (the so-called  recovery variable). The
functions $F(x)$ and $H(x-d)$ are of the form
\begin{eqnarray}\label{eq:nonLn}
  F(x) &=& \begin{cases}
             -m_0x, \text{ if } x \leq J_{min} \\
             m_1(x-a), \text{ if } J_{min} < x < J_{max} \\
             -m_0(x-1), \text{ if } x \geq J_{max}
           \end{cases} \\
  H(x) &=& \begin{cases}
             1, \text{ if } x \geq 0 \\
             0, \text{ if } x < 0
           \end{cases}
\end{eqnarray}
where
\begin{eqnarray*}
     J_{min} = \frac{a m_1}{m_0+m_1}, \quad
         J_{max} = \frac{m_0 + a m_1}{m_0+m_1}, \quad
         m_0,m_1 > 0.
\end{eqnarray*}
The parameter $\varepsilon$ $( \varepsilon > 0 )$ defines the time
scale of recovery variable, the parameter $J$ is a constant
external stimulus, the parameters $\beta$ $(\beta > 0)$ and $d$
$(d>0$) control the threshold property of the bursting oscillations.
Here we have chosen this linear piece-wise approximation of $F(x)$
in order to obtain a simple hyperbolic map for chaotic
spiking-bursting activity. However, any cubic function can be also
used. The map $f$ is discontinuous map and $x=d$ is the
discontinuity line of $f$. We consider only those trajectories
(orbits) which do not fall within a discontinuity set
$D=\bigcup_{i=0}^\infty f^{-i}L$, where $L$ is the union of points
of discontinuity of $f$ and its derivative $Df$. Besides, we assume that $m_0<1$,
then
\begin{equation*}
     \det \frac{ \partial(\overline{x},\overline{y})}
         { \partial(x,y) } = 1 + F'(x) +\varepsilon > 0
\end{equation*}
for any $\varepsilon \geq 0$ and the map $f$ is one to one. We
restrict consideration of the dynamics of the map $f$ to the
following parameter region
\begin{equation} \label{eq:condD}
     0<J<d, \quad J_{min} < d < J_{max}, \quad m_0 < 1.
\end{equation}
Note that under such conditions we have $F'(d)>0$. This condition
is very important for forming chaotic behavior of the map  $f$ as
we shall see bellow. For convenience, we rewrite the map $f$ in
the following form
\begin{equation*}
     f = \begin{cases}
         f_1, \text{ if } x<d \\
         f_2, \text{ if } x \geq d
     \end{cases}
\end{equation*}
where $f_1$, $f_2$ are the maps
\begin{eqnarray*}
     f_1&:&(x,y)\rightarrow (x+F(x)-y, \quad y + \varepsilon (x-J)) \\
     f_2&:&(x,y)\rightarrow (x+F(x)-y-\beta, \quad y + \varepsilon
     (x-J)).
\end{eqnarray*}

\section{ \label{sec:OneDDyn} One-dimensional dynamics of the model \protect\\}
Let us start with the dynamics of the map $f$ when the parameter
$\varepsilon=0$. In this case the map $f$ is reduced to a one
dimensional map:
\begin{equation}\label{eq:oneDMap}
     \overline{x} = x +F(x) -y_0 -\beta H(x-d) = g(x)
\end{equation}
where $y_0$ is a constant and it plays the role of a new
parameter. The map (\ref{eq:oneDMap}) can be rewritten as
\begin{equation} \label{eq:oneDMapFl}
     g(x) = \begin{cases}
         g_1(x) \equiv (1-m_0)x-y_0, \text{ if } x \leq J_{min} \\
         g_2(x) \equiv  q x - y_0 -a m_1,\text{ if }
             J_{min} < x< d \\
         g_3(x) \equiv  q x - y_0 -a m_1 - \beta, \text{ if }
             d \leq x \leq J_{max} \\
         g_4(x) \equiv  (1-m_0) x - y_0 + m_0 - \beta, \text{ if }
             x > J_{max},
         \end{cases}
\end{equation}
where $q=1+m_1$.
\par
\begin{figure}[t]
   \includegraphics[width=0.95\textwidth]{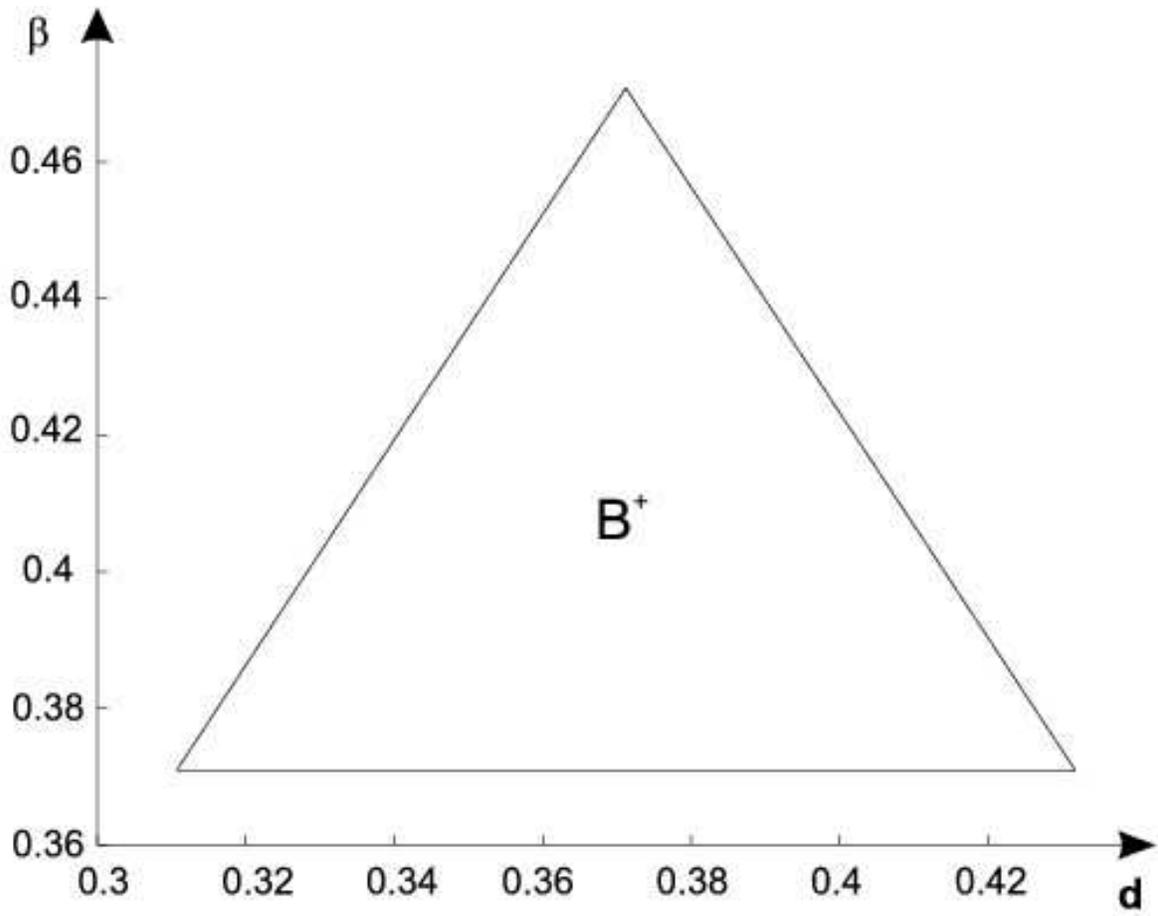}\\
   \caption{The parameter region $B^+$ defined by inequalities
     (\ref{eq:Cnd1DBeta}) and (\ref{eq:CndLrnc1})
     ($m_0=0.0864, m_1 = 0.65, a = 0.2$)}\label{pic:BD_dBeta}
\end{figure}
\begin{figure}[t]
   \includegraphics[width=0.95\textwidth]{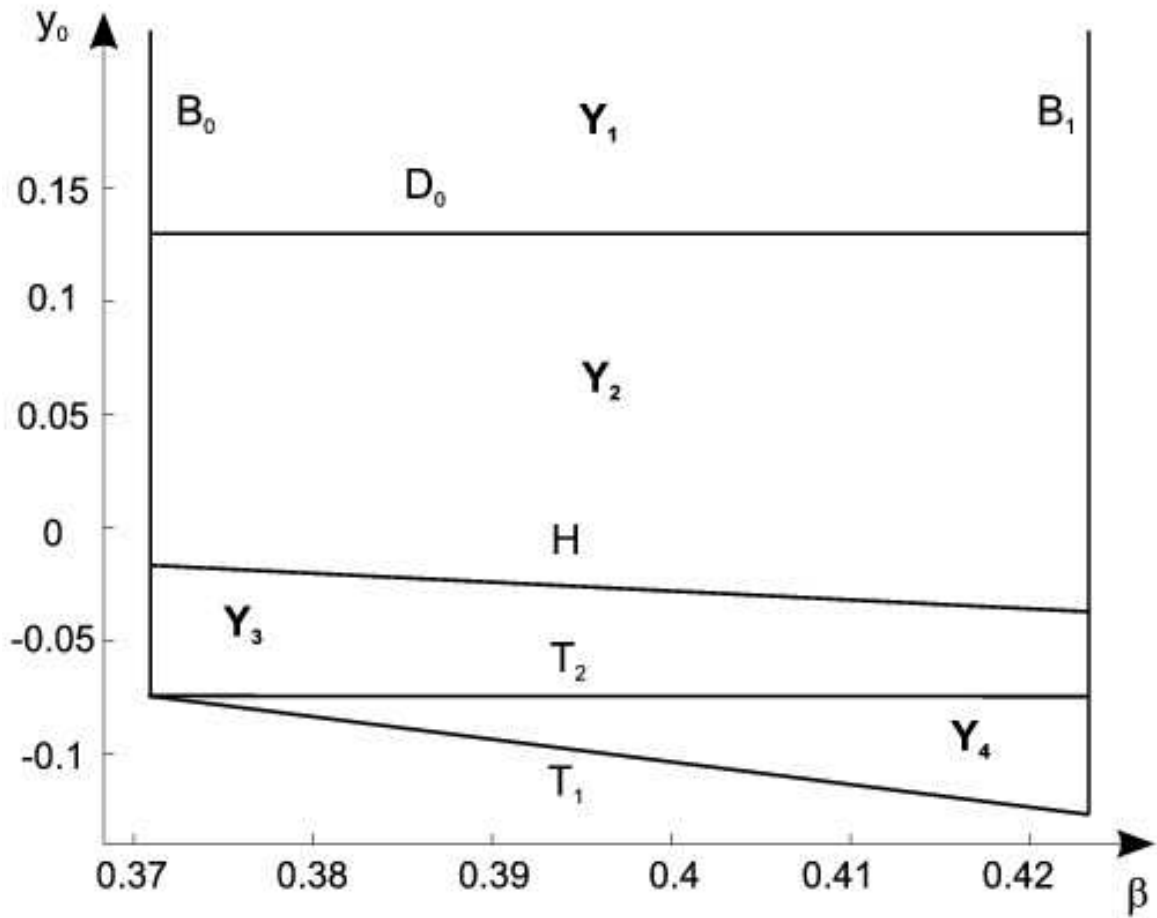}\\
   \caption{The $Y$ region on the $(\beta, y_0)$ plane with
     (\ref{eq:Cnd1DBeta}) and (\ref{eq:Cnd1DY0})
     ($m_0=0.0864, m_1 = 0.65, a = 0.2, d = 0.4 $)}\label{pic:BD_betaY0}
\end{figure}
\begin{figure}[t]
   \includegraphics[width=0.75\textwidth]{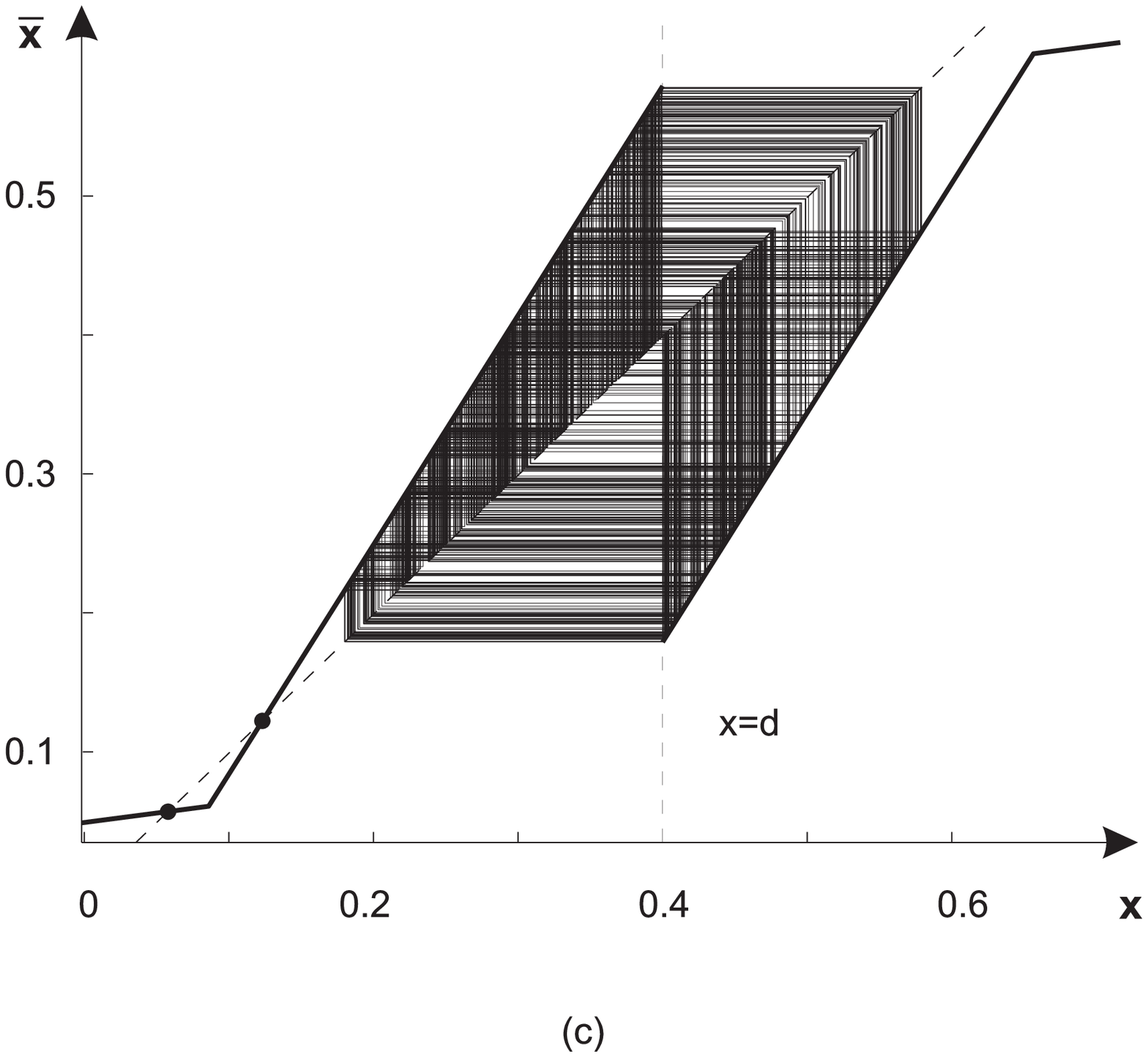}
   \includegraphics[width=0.75\textwidth]{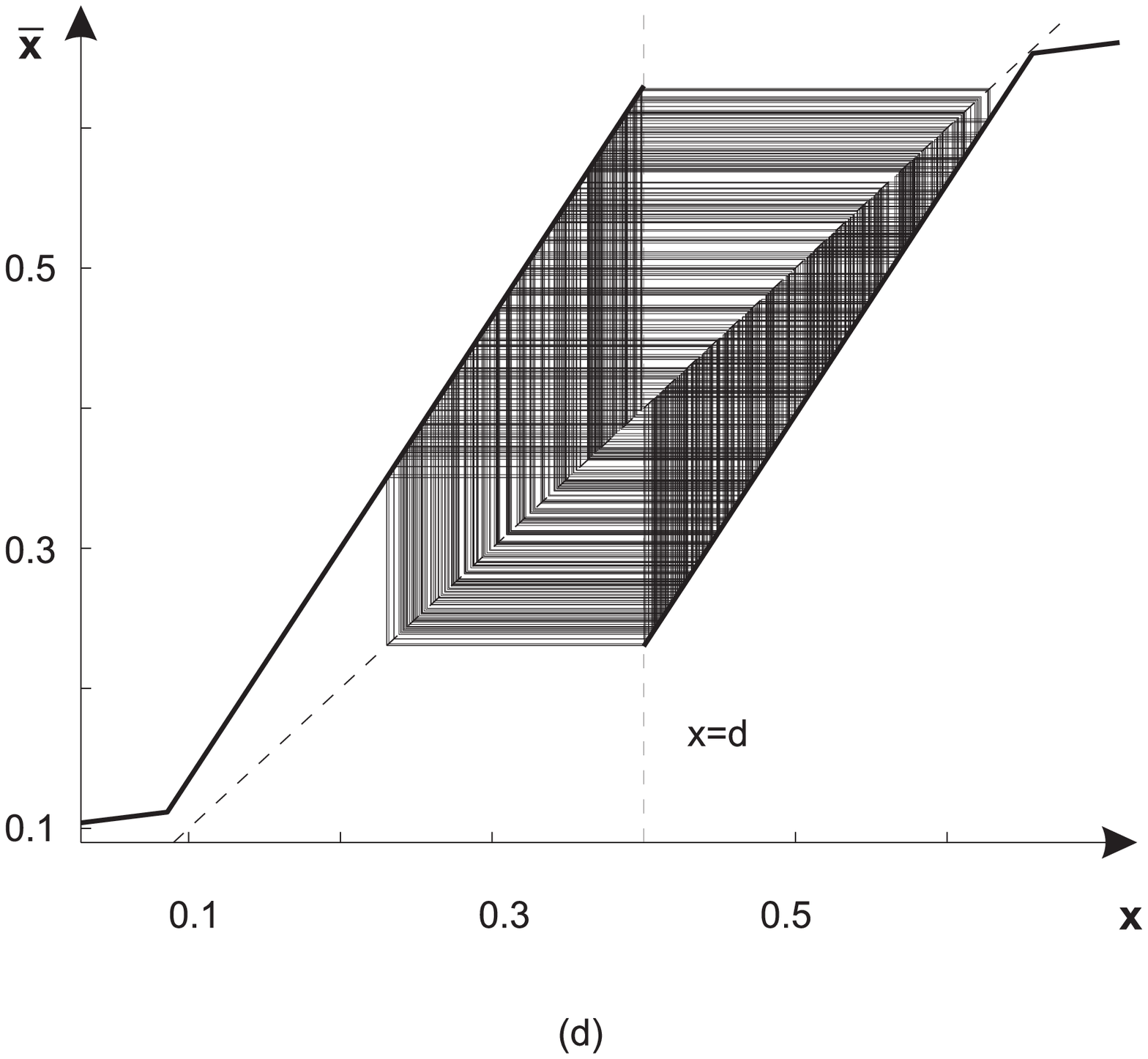} \\
   \caption{The dynamics of the map $g$ with
     $m_0=0.0864, m_1 = 0.65, a = 0.2, d = 0.4, \beta = 0.4$ for
     regions: (a) - $Y_1$, $y_0=0.15$, (b) - $Y_2$, $y_0=0.05$,
     (c) - $Y_4$, $y_0=-0.05$, (d) - $Y_3$, $y_0=-0.1$.}\label{pic:KL}
\end{figure}

Let us fix the parameters $a$, $d$, $m_0$, $m_1$ and consider the
dynamics of the map (\ref{eq:oneDMapFl}) in the parameter plane
$(\beta, y_0)$. We restrict our study  of the map
$f$ to the following parameter region
\begin{equation} \label{eq:Cnd1DBeta}
     \beta > \beta_0
\end{equation}
\begin{equation} \label{eq:Cnd1DY0}
     y_0 > F( J_{max} ) - \beta
\end{equation}
where $\beta_0 = F(J_{max}) - F(J_{min})$. These conditions allow
to obtain interesting properties of the map (\ref{eq:stpMp}). Let
us find the conditions on the parameter values for which the map
$f$ acts like a Lorenz-type map \cite{b:Afraimovich}. For that we
require that (see Fig. \ref{pic:KL})
\begin{equation} \label{eq:CndLrnc0}
     \lim_{x\nearrow d} g_2(x) < g_3(J_{max}), \qquad
         \lim_{x \searrow d} g_3(x) > g_2(J_{min}).
\end{equation}
It follows from (\ref{eq:CndLrnc0}) the following condition on the
parameter $\beta$:
\begin{equation} \label{eq:CndLrnc1}
     \beta < \beta_1,
\end{equation}
where
\begin{equation*}
     \beta_1 = min \left\{ q(J_{max}-d), q(d-J_{min}) \right\}.
\end{equation*}
The inequalities (\ref{eq:Cnd1DBeta}) and (\ref{eq:CndLrnc1})
define on the $(d,\beta)$ plane the region $B^+$ (see
Fig.\ref{pic:BD_dBeta}). Let us take the parameters $d$ and
$\beta$ inside the $B^+$ region, and let us consider the $(\beta,
y_0)$ plane. In this plane the inequalities (\ref{eq:Cnd1DBeta}),
(\ref{eq:Cnd1DY0}) and (\ref{eq:CndLrnc1}) are satisfied
simultaneously in region $Y$. In this plane the boundary of $Y$
consists of the three lines (Fig. \ref{pic:BD_betaY0})
\begin{eqnarray*}
     B_0&=&\left\{ (y_0,\beta):
         \beta = \beta_0,\quad y_0 \geq F(J_{min}) \right \} \\
     B_1&=&\left\{ (y_0,\beta):
         \beta = \beta_1,\quad y_0 \geq F(J_{max})-\beta_1 \right \} \\
     T_1&=&\left\{ (y_0,\beta):
         y_0 =  F(J_{max})-\beta,\quad
         \beta_0 \leq \beta \leq \beta_1 \right \}.
\end{eqnarray*}
Consider the dynamics of the map $g$ for $(y_0,\beta) \in Y$. This
region is separated on four subregions by the bifurcation lines
\begin{eqnarray*}
     D_0&=&\left\{ (y_0,\beta):
         y_0 = F( d ),\quad
         \beta_0 \leq \beta \leq \beta_1 \right \} \\
     H&=&\left\{ (y_0,\beta):
         y_0 = F( d ) - \frac{q-1}{q}\beta,\quad
         \beta_0 \leq \beta \leq \beta_1\right \} \\
     T_2&=&\left\{ (y_0,\beta):
         y_0 =  F(J_{min}),\quad
         \beta_0 \leq \beta \leq \beta_1 \right \},
\end{eqnarray*}
corresponding to different dynamics of the map $g$. The line $D_0$
coincides with appearance of an unstable fixed point $x=a+y_0/m_1$
through crossing of the discontinuity point $x=d$. Line $T_2$
corresponds to the fold (tangent) bifurcation of the fixed point
$x=J_{min}$ (see Fig. \ref{pic:KL}(a,d)). Line $H$ corresponds to
the condition
\begin{equation*}
     \lim_{x \searrow d} g_3(x) = a + y_0/m_1.
\end{equation*}
Note that for $(y_0,\beta) \in H$ there exists a bifurcation
corresponding to appearance of homoclinic orbit
\cite{b:AfraimovichShilnikov} to the unstable fixed point. The
dynamics of the map $g$ corresponding to subregions $Y_i(i=1,..4)$
is shown in Fig.\ref{pic:KL}. If $(\beta, y_0) \in Y_1 \bigcup
Y_2$ the trajectories of the map $g$ tend to stable fixed point
$x=-y_0/m_0$ for any initial conditions different from an unstable
fixed point (Fig.\ref{pic:KL} (a), (b)). If $(\beta, y_0) \in Y_3
\bigcup Y_4$ the map $f$ has invariant interval $I=\{x:\quad b< x
< c\}$, where
\begin{eqnarray} \label{eq:invBrd}
     b &=& q d - y_0 - a m_1 - \beta \\
     c &=& q d -y_0 - a m_1. \nonumber
\end{eqnarray}
For parameters $(\beta, y_0) \in Y_3$ the map $g$ exhibits
bistable property, that is there exists two attractors, one is a
stable fixed point and the second is an invariant set of the
interval $I$ whose basins of attraction are separated by an
unstable fixed point (Fig.\ref{pic:KL}(c)). For $(\beta, y_0) \in
Y_4$ there exists the interval $I$ (Fig.\ref{pic:KL} (d)) which
attract all trajectories of the map $g$.
\par
Check that the map $g$ on the $I$ acts like a Lorenz-type map. The
map $g$ will be a Lorenz-type if \cite{b:Afraimovich}
\begin{enumerate}
     \item \label{lst:cndLrnc1} the derivative $g'(x)>0$ for any
         $x \in I \setminus \{ d\}$;
     \item \label{lst:cndLrnc2} the set of preimages of the point of
discontinuity,
         $D =\bigcup \limits_{n \geq 0} g^{-n}(d)$ is dense in $I$;
     \item \label{lst:cndLrnc3} $\lim \limits_{x \searrow d}
         g(x) = b, \quad \lim \limits_{x \nearrow d} g(x) = c$.
\end{enumerate}
One can see that (\ref{lst:cndLrnc1}) and (\ref{lst:cndLrnc3}) are
satisfied. According to \cite{b:Afraimovich} the condition
(\ref{lst:cndLrnc2}) is satisfied if
\begin{equation} \label{eq:CndLrncIII}
     g'(x) \geq q > 1, \quad x \in I \setminus \{d \}.
\end{equation}
For the map $g$ on the interval $I$ we have $q=1+m_1$ and inequality
(\ref{eq:CndLrncIII}) is obviously satisfied. Therefore the map
$g$ on the interval $I$ acts like a Lorenz-type map. The possible structure
of the invariant set of interval $I$ is controlled by value $q$.
\par
Let us find conditions under which the map $g$ is strongly
transitive. Recall \cite{b:Afraimovich} that a Lorenz-type map $g$
is strongly transitive if for any subinterval $I_0 \subset I
\setminus \{d\}$ there is $k \geq 0 $ such that
\begin{equation*}
     \bigcup_{i=0}^k f^i I_0 \supset Int I.
\end{equation*}
Under the condition (\ref{eq:CndLrncIII}) the sufficient condition
for strong transitivity on the interval $I$ are (
\cite{b:Afraimovich})
\begin{equation} \label{eq:CndST}
     \min \left \{ q^{n_1+1}, q^{n_2+1} \right \} > 2
\end{equation}
where $n_1, n_2 \in \mathbb{Z}_+$ are such that they  satisfy the
following conditions
\begin{equation} \label{eq:CndN1}
     g_2(b), \ldots, g_2^{n_1}(b) \in [ b,d), g_2^{n_1+1}(b) \in (d,c]
\end{equation}
\begin{equation}\label{eq:CndN2}
     g_3(c), \ldots, g_3^{n_2}(c) \in ( d,c ], g_3^{n_2+1}(c) \in [b,d).
\end{equation}
Now let us find condition for the parameter  values of the map $g$
under which $n_1=n_2 = 1$. Consider the condition (\ref{eq:CndN1}).
Let us take $n_1=k$, where $k=1,2,\ldots$. It is clear that
(\ref{eq:CndN1}) holds if the parameter $y_0$ satisfies the
following conditions
\begin{equation} \label{eq:Cnd:StrTransN1RngY0}
     \begin{cases}
         y_0 \geq F(d) - \beta q^k (q-1)/(q^{k+1}-1) \\
         y_0 < F(d) -\beta q^{k+1} (q-1)/(q^{k+2}-1).
     \end{cases}
\end{equation}
Let us require that (\ref{eq:Cnd:StrTransN1RngY0}) for $k=1$ is
satisfied for
\begin{equation} \label{eq:CndStrTransPars}
     (\beta,d) \in B^+, \quad y_0 \in Y_3 \bigcup Y_4.
\end{equation}
 From inequalities (\ref{eq:Cnd1DBeta}), (\ref{eq:CndLrnc1}) and
the definitions of the boundaries $T_1$ and $H$, it follows that
this requirement holds if
\begin{equation} \label{eq:CndStrTransQ}
     q \geq \frac{1+\sqrt{5}}{2}.
\end{equation}
Similarly, for $n_2=k$ we get
\begin{equation} \label{eq:Cnd:StrTransN2RngY0}
     \begin{cases}
         y_0 > F(d) - \beta (q^{k+1}-1)/(q^{k+2}-1) \\
         y_0\leq F(d) -\beta (q^k-1)/(q^{k+1}-1).
     \end{cases}
\end{equation}
By the same argument as indicated above we  obtain that for
$n_2=1$ inequalities (\ref{eq:Cnd:StrTransN2RngY0}) hold if the
conditions (\ref{eq:CndN2}) are satisfied. For example,
let us fix $q=1,65$, that is $m_1=0.65$. In this case the map
$g|_I$ is strongly transitive and therefore it follows from the
theorem 3.1.1. of \cite{b:Afraimovich} that the periodic points
are dense in $I$. We note that all of these periodic points are
unstable $(q>1)$ and $I$ is a chaotic attractor. Fig.\ref{pic:KL}
(c),(d) illustrates the dynamics of the map $g$ on the interval
$I$ for regions $Y_3$ and $Y_4$ respectively.

\section{ \label{sec:TwoDDyn} Relaxation two-dimensional dynamics
of the model and spiking-bursting oscillations \protect\\} In this
section consider the case $\varepsilon << 1$ and $J > J_{min}$.
This case corresponds to instability of the unique fixed point
$O(x=J, y = F(J)$. Since parameter $\varepsilon$ is sufficiently
small, the dynamic of the map $f$ is a relaxation \cite{b:Arnold}
similarly by to the case of ODE (\ref{eq:ModelFHN}) . The
distinctive characteristic of these systems is two time and
velocity scales, so-called "fast" and "slow" motions. Basically
fast motions are provided by "frozen" system in which slow
variables are regarded as a parameters, and it is assumed that
small parameter of the system equals to zero. Slow motions with
size of order of the small parameter are given by evolution of
"frozen" variable. In case of the map $f$, $x$ is the fast variable
and $y$  is the slow one. Let us study the fast and slow motions in
our system.

\subsection{ \label{sec:FastAnsSlowMt} Fast and slow motions }
The fast motions of the model (\ref{eq:stpMp}) is approximately
described by the map (\ref{eq:oneDMap}). As indicated above, the
dynamics of the map (\ref{eq:oneDMap}) can be both, regular and
chaotic according to the parameter value (Fig.\ref{pic:KL}).
Consider now under conditions (\ref{eq:Cnd1DBeta}),
(\ref{eq:CndLrnc1}) slow motions of the map $f$ on the phase plane
$(x,y)$ in the region separated by the following inequalities
\begin{equation}\label{eq:cndSMXY}
     x<J_{min}, \quad  y > F(J_{max}) - \beta
\end{equation}
In the case $\varepsilon << 1$ the motions of the map $f$ have
slow features within thin layer $M_1^s(\varepsilon)$ (thickness is
of the order $\varepsilon^\alpha$, $0<\alpha < 1$) \cite{b:Arnold}
near invariant line
\begin{equation*}
     W_1^S(\varepsilon) = \left \{ (x,y):y=-k_0x-b_0, \quad
         x <J_{min} \right \},
\end{equation*}
where
\begin{equation}\label{eq:smK0B0}
     k_0=\frac{m_0}{2} + \sqrt{ \frac{m^2_0}{4} - \varepsilon },
         \quad b_0=\frac{\varepsilon J (1-m_0+k_0)}{k_0-\varepsilon}.
\end{equation}
Directly from the map $f$ it can be obtained that
$W_1^S(\varepsilon)$ is invariant line not only for $\varepsilon
\rightarrow 0$ but for $\varepsilon \leq m_0^2/4$ also. One can
see that the dynamics on the line $W_1^S(\varepsilon)$ is defined
by one-dimensional linear map
\begin{equation} \label{eq:smOneDMap}
     \overline{x} = (1-m_0+k_0)x+b_0.
\end{equation}
It is clear that the map (\ref{eq:smOneDMap}) has stable fixed
point $x=J$. Therefore for $J>J_{min}$ the trajectories on
$W_1^S(\varepsilon)$ with initial conditions $x<J_{min}$ moves to
the line $x=J_{min}$. All trajectories from layer
$M_1^s(\varepsilon)$ behave in the same way.
\par
Let us now consider the stability of the slow motions from
$M_1^s(\varepsilon)$ relatively  to the fast ones. Since in the case
$\varepsilon = 0$ each point of the $W^S(0)$ is stable fixed point
of the fast map (\ref{eq:oneDMap}) then invariant curve
$W_1^S(\varepsilon)$ is stable with respect to the fast motions.

\subsection{ \label{sec:Chaos} Relaxation chaotic
dynamics \protect\\}

It is follows from the previous description that when
$\varepsilon$ is small enough, the structure of the partition 
of the phase $(x,y)$-plane into trajectories doesn't significantly
change with respect to case of equations  (\ref{eq:oneDMap}),
(\ref{eq:smOneDMap}). The trajectories of the map $f$ are close to
the trajectories of (\ref{eq:smOneDMap}) within the layer of the
slow motions near $W_1^S(\varepsilon)$ and to the trajectories of
(\ref{eq:oneDMap}) outside these layer. Therefore, the motions of
the map $f$ are also formed by the slow-fast trajectories.
\par
\begin{figure}[t]
   \includegraphics[width=0.75\textwidth]{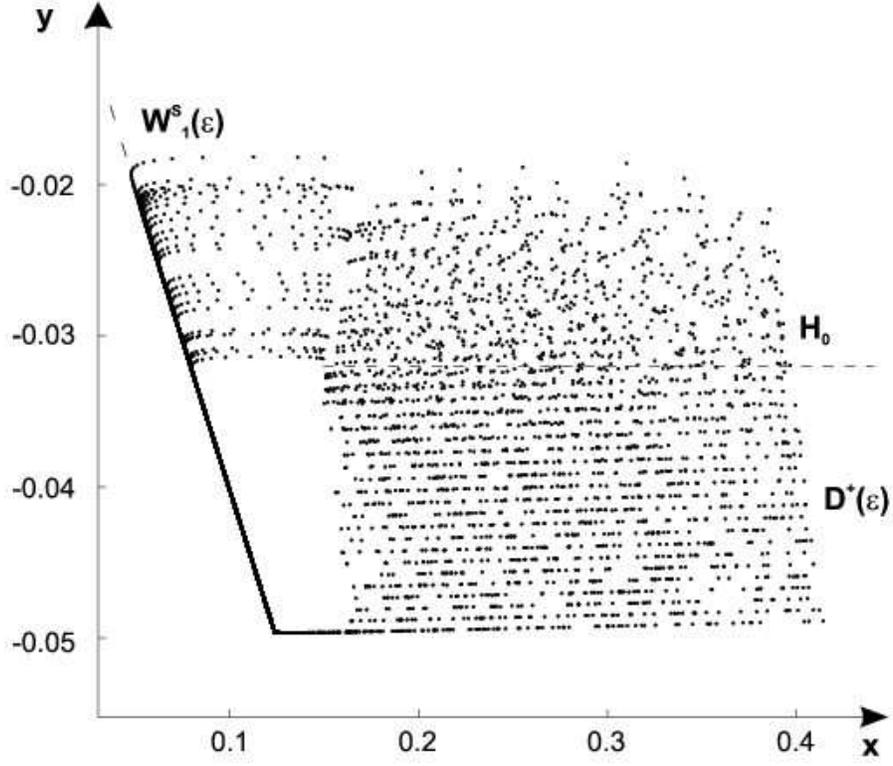} \\
   \includegraphics[width=0.75\textwidth]{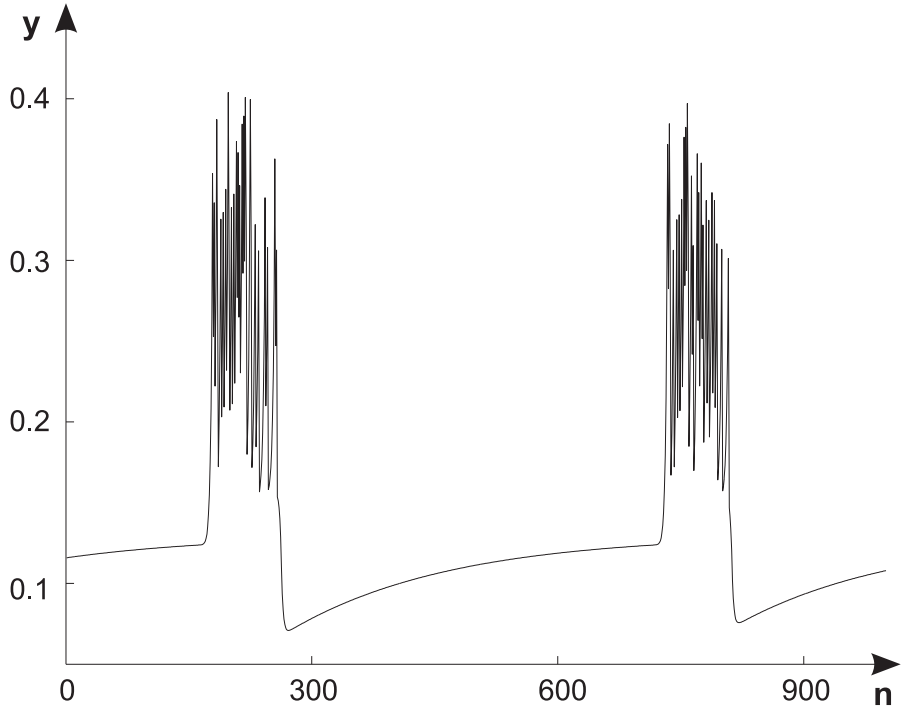} \\
     \caption{ (a) The chaotic attractor $A$ on the phase plane
     $(x,y)$; (b) Waveform of relaxation spike-bursting
     oscillations generated by the map $f$.
     Parameter values: $J = 0.13, m_0 = 0.4, m_1 = 0.65, a = 0.2, d = 0.3,
     \beta = 0.25, \varepsilon = 0.002 $.
      }\label{pic:ChAtt}
\end{figure}

Let the initial conditions of $f$ belong to neighborhood
$M_1^S(\varepsilon)$ of the invariant curve $W_1^S(\varepsilon)$.
Any of these trajectories moves within the layer of the slow
motions down to the neighborhood of the critical point $C$ : $x
\approx J_{min}, y \approx F(J_{min})$, and continue their motions
according to the fast motions (see Fig. \ref{pic:KL}(a)), along $y
\approx y_0 = F(J_{min})$. Since $y_0 \in Y_4$ the trajectory of
the map $g$ with initial condition $x \approx J_{min}$ tends to
invariant interval $I$ (See Fig. \ref{pic:KL} (d)). Therefore the
fast motions of the map $f$ with initial conditions $C$  falls
into some region $D^+(\varepsilon)$ (see Fig \ref{pic:ChAtt}),
$D^+(\varepsilon) \rightarrow D^+(0)$ if $\varepsilon \rightarrow
0$, where $D(0)$ is the parallelogram:
\begin{eqnarray}
     D^+(0) = \{ (x,y) &:& \nonumber \\
         q d - y -a m_1 - \beta \leq  &x& \leq q d - y - am_1, \\
      F(J_{max}) - \beta \leq &y& \leq F(d) - \beta(q-1)/q  \nonumber \}
\end{eqnarray}
In other words the region $D^+(0)$ is one parametrical family
$y_0$ - indexed of invariant intervals. As $I$ is attractor, then
$D^+(\varepsilon)$ is also two-dimensional attracting region.
Since the map $g$ has interval $I$ for $y_0 \in Y_3 \bigcup Y_4$,
then a trajectory involving the map $f$ belongs to the region
$D^+(\varepsilon)$ as long as its variable $y$ do not culminate
approximately to the value corresponding to the line $H_0$ (Fig.
\ref{pic:BD_betaY0}). At the same time variable $y$ is slowly
increasing for $(x,y) \in D^+(\varepsilon)$ as $D^+(\varepsilon)
\in \{ x>J \}$. Thus, within the region $D^+(\varepsilon)$ the
variable $y$ continues to increase and variable $x$ evolution is
close to chaotic trajectory of the map $g$.
\par
Over line $H (y_0 \in  Y_2)$  the map $g$ has stable fixed point
which attracts all trajectories (see Fig.\ref{pic:KL} (b)). Hence
if the magnitude of the variable $y$ becomes about $H$ then
trajectory of the map $f$  returns into neighborhood of
$M_1^s(\varepsilon)$. Then the process is repeated. As a result of
these slow-fast motions the attractor $A$ of the system $f$
  $(x,y)$ phase plane appears as in  (Fig.\ref{pic:ChAtt}(a)).
\par
To characterize the complexity of the attractor A we calculated
numerically its fractal dimension $d_f(A)$. At appears that
$d_f(A)$ takes non-integer values between 1.35 and 1.9
(Fig.\ref{pic:ChAtt}(b)). Therefore $A$ is chaotic attractor. For
the parameter values from Fig. \ref{pic:ChAtt} (b) maximum of the
fractal dimension $d_f(A) = 1.8287$ is accomplished then $J =
0.2661$.

\section{ \label{sec:InvR} Invariant region,
chaotic attractor and spiking-bursting oscillations\protect\\}

Let us prove that the system (\ref{eq:stpMp}) has an attractor $A$
for different values $\varepsilon$ and let us find conditions under
which the map
  $f$ has an
invariant region. To do that, we construct some ring-like region
$S$. Denote by $\Gamma$ the outer boundary and by $\gamma$ the
inner boundary of the $S$. The $S$ is an invariant region if from
the conditions $(x,y) \in S$ and $(x,y) \not \in D$ follows that
$(\overline{x}, \overline{y}) \in S$. Its should be fulfilled if
\begin{enumerate}
     \item \label{lst:cndInv1} the vector field of the map $f$ at the
         boundary $\Gamma$ and $\gamma$ is oriented inwards to $S$;
     \item \label{lst:cndInv2} the images $f_i(\Gamma),f_i(\gamma), f_i(d),
         i=1,2$ of the boundaries $\Gamma$, $\gamma$ and the
         discontinuity line $d$ belong to $S$.
\end{enumerate}
We construct boundaries $\Gamma$ and $\gamma$ in the form of some
polygons. Taking into account the condition (\ref{lst:cndInv1})
and analyzing the vector field of the map $f$ at the lines with
uncertain slope we have found the shape of $\Gamma$ and $\gamma$
(see Fig. \ref{pic:PM_InvRg} (a)). The equations of the boundaries
of $\Gamma$ and $\gamma$  are presented in the Appendix. Analysis of the
position of the images $f_i(\Gamma), f_1(\gamma), f_i(d), (i=1,2)$
on the phase plane $(x,y)$ show that the condition
(\ref{lst:cndInv2}) holds if $(\beta,d) \in B^+$ (see section
\ref{sec:OneDDyn}) and inequalities

\begin{equation*}
     -\frac{ \sqrt{ \varepsilon }  }{m_0}(J - J_{min})
         - m_0 J - k_1(d-J) > F(J_{max}) - \beta
\end{equation*}

\begin{equation*}
     \sqrt{\varepsilon} < \min \left\{
         \frac{B}{2(d-J)},
         \frac{m_0\left[B - m_1(J_{max}-d) \right] }{J - J_{min}} \right\}
\end{equation*}

\begin{figure}[t]
   \includegraphics[width=0.75\textwidth]{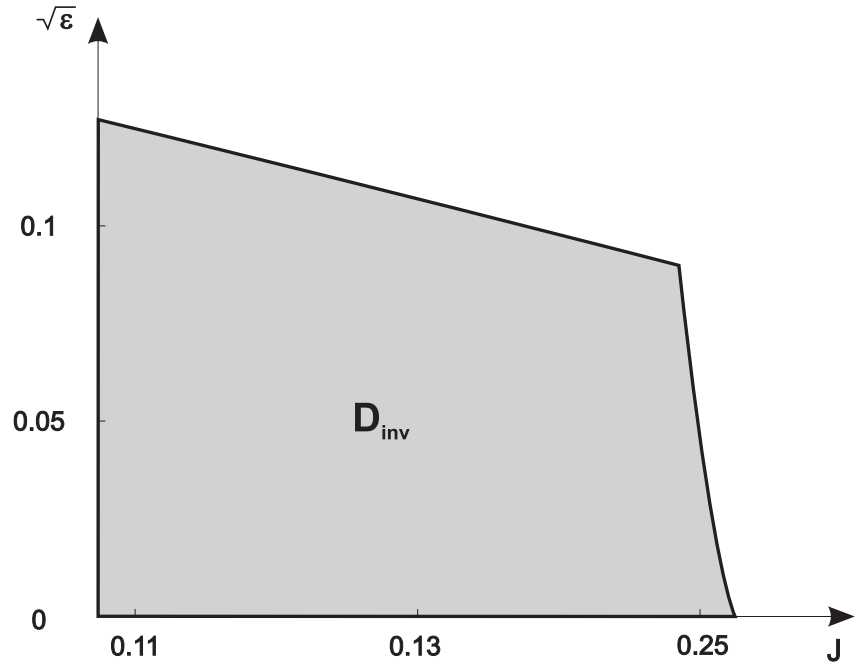} \\
   \includegraphics[width=0.75\textwidth]{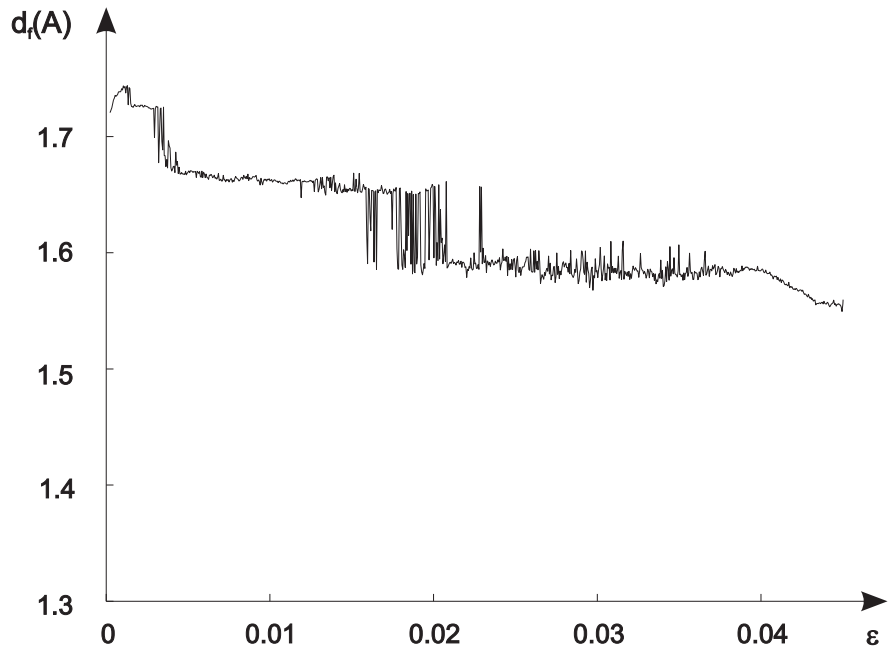} \\
   \caption{ (a) Parameter region $D_{inv}$ on the parameter plane
$(J,\varepsilon)$,
   parameter values: $m_0 = 0.5, m_1 = 0.65, a = 0.2, d = 0.34,
     \beta = 0.31$. (b) Fractal dimension $d_f(A)$ of the attractor $A$
     versus parameter $\varepsilon$, $J = 0.15$ }\label{pic:BdInvRg}
\end{figure}

\begin{figure}[t]
   \includegraphics[width=0.7\textwidth]{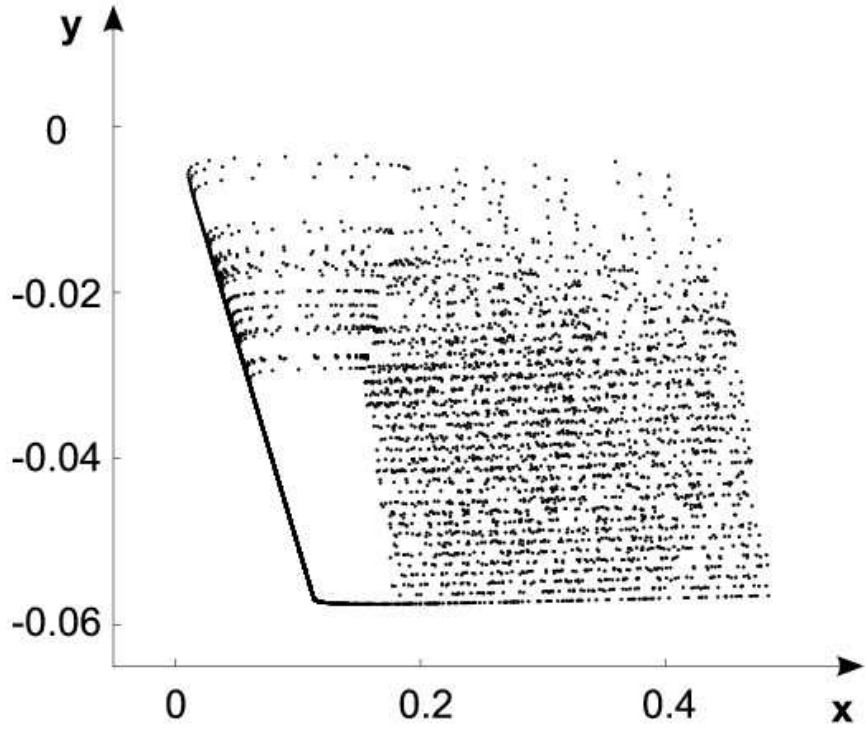} \\
   \includegraphics[width=0.7\textwidth]{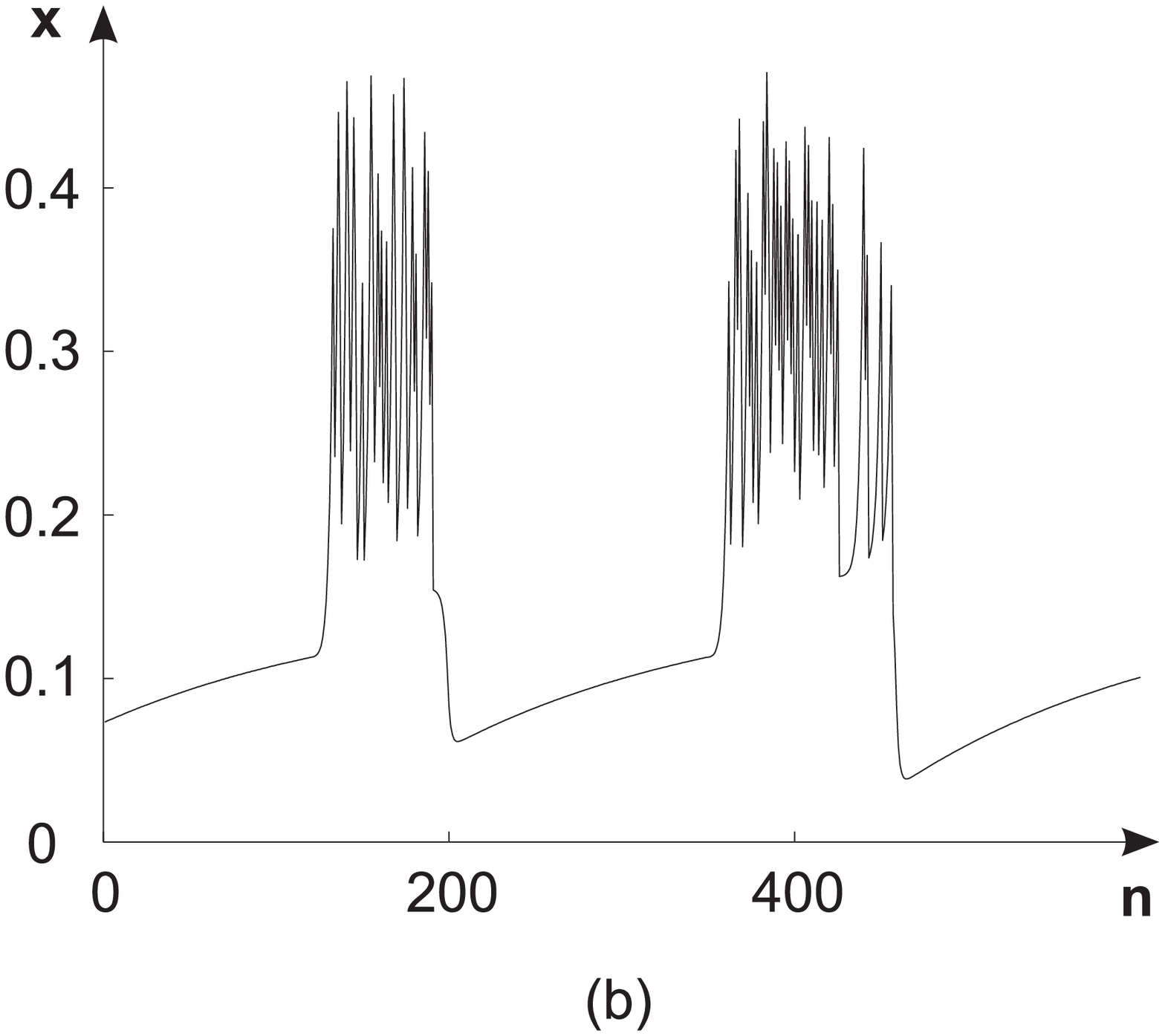}\\
   \caption{ (a) Chaotic attractor $A$ on the phase plane $(x,y)$.
     (b) Spike-bursting oscillations generated by
     the map $f$. Parameters value $m_0 = 0.5, m_1 = 0.65, a = 0.2, d = 0.34,
     \beta = 0.31, J=0.15, \varepsilon = 0.004 $ } \label{pic:PM_Bursts}
\end{figure}

\begin{widetext}
     \begin{equation*}
         \frac{\sqrt{\varepsilon}}{m_0}(J-J_{min}) + m_1(J_{max}-d)
             + m_1(J_{max}-d) < \frac{B}{2}
             + \sqrt{ \frac{B^2}{4} - \varepsilon (d-J)^2 },
             \text{ if } \sqrt{\varepsilon} < \frac{B}{2(d-J)}
     \end{equation*}
\end{widetext}

\begin{equation} \label{eq:CndInvEpsJ}
     \sqrt{\varepsilon} <
         \frac{\beta - m_1 (J_{max}-d)}{2(J_{max}-d)},
         \quad J_{min} < J < \frac{d(1+m_1)-am_1 - \beta}{1-m_0}
\end{equation}

\begin{equation*}
     \varepsilon(d-J)
         - \frac{(1+m_1)(J-J_{min})}{m_1}\sqrt{\varepsilon}
         + (1+m_1)(d-J) - \beta > 0,
\end{equation*}

\begin{equation*}
     d + F(d) + m_0J - J_{max}
         + \frac{\sqrt{\varepsilon}}{m_0} ( d - J_{min})
         + k_1(d - J ) < 0
\end{equation*}

\begin{equation*}
     \sqrt{\varepsilon} < \min \left \{ \frac{m_0}{2}, \quad
         \frac{m_1}{2}, \quad \frac{m_1(d-J)}{J-J_{min}}  \right \}
\end{equation*}
are satisfied (the parameters $k_1$, $B$ and $J_0$ have been
introduced in Appendix). Fig.\ref{pic:PM_InvRg}(b),(c) illustrates
the transformation of $S$ by the action of the map $f$ under
conditions (\ref{eq:CndInvEpsJ}). The inequality
(\ref{eq:CndInvEpsJ}) determine the parameter region $D_{inv}$ in
the parameter plane $(J,\varepsilon)$ (Fig. \ref{pic:BdInvRg}
(a)). Since
\begin{eqnarray} \label{eq:CndInvRg}
     \overline{y}  < y, \text{ for }
         (x,y) \in S \bigcap \{ x <J \}\\
     \overline{y}  > y, \text{ for }
         (x,y) \in S \bigcap \{ x >J \}
\end{eqnarray}
then the trajectories with initial conditions $(x,y) \in S$
execute rotation motion around the fixed point $O$ forming some attractors $A$. We calculated numerically
fractal dimensional  $d_f(A)$ (Fig.\ref{pic:BdInvRg}(b)) in terms
of $\varepsilon$. Its shows that $A$ is chaotic attractor. The
possible structure of the attractor $A$ in the phase plane is
shown in Fig.\ref{pic:PM_Bursts}(a). Fig. \ref{pic:PM_Bursts}(b)
illustrates time evolution of the variable $x$ corresponding to
chaotic attractor $A$. It shows chaotic spiking-bursting neural
activity. Fig. \ref{pic:BdInvRg} (b) shows that fractal dimension
$d_f(A)$, on average, tends to decrease with increasing
$\varepsilon$. There is a critical value, $\varepsilon = 0.0461$,
for which fractal dimension has a minimal value $d_f(A)=1.5114$.
The mechanism of this decreasing can be accounted for by the
different types of the dynamics of the variable $y$ for different
$\varepsilon$. As the parameter $\varepsilon$ increases, the
velocity of the variable $y$ is expected to climb. Therefore "life
time" of the trajectories in the strip corresponding to Lorenz-map
dynamics is reduced. As a result, the chaotical motions are
reduced.

\section{ \label{sec:OtherModes} The gallery of the other attractors
         and the regimes of neural activity \protect\\}

At previous sections it was shown that system (\ref{eq:stpMp})
allows to simulate spiking-bursting behavior of the neuron. Here
we show that other regimes of the neural activity (phasic spiking
and burstings threshold excitation, subthreshold oscillation,
tonic spiking and chaotic spike generation) can be obtained by
using the map $f$ also. To do that we neglect the first inequality
in (\ref{eq:condD}), inequality (\ref{eq:Cnd1DBeta}) and condition
$y > F(J_{max})-\beta$.

\subsection{ \label{sec:OtherExcit} The generation of phasic spikes
and bursting} 
\begin{figure}[t]
   \includegraphics[width=0.7\textwidth]{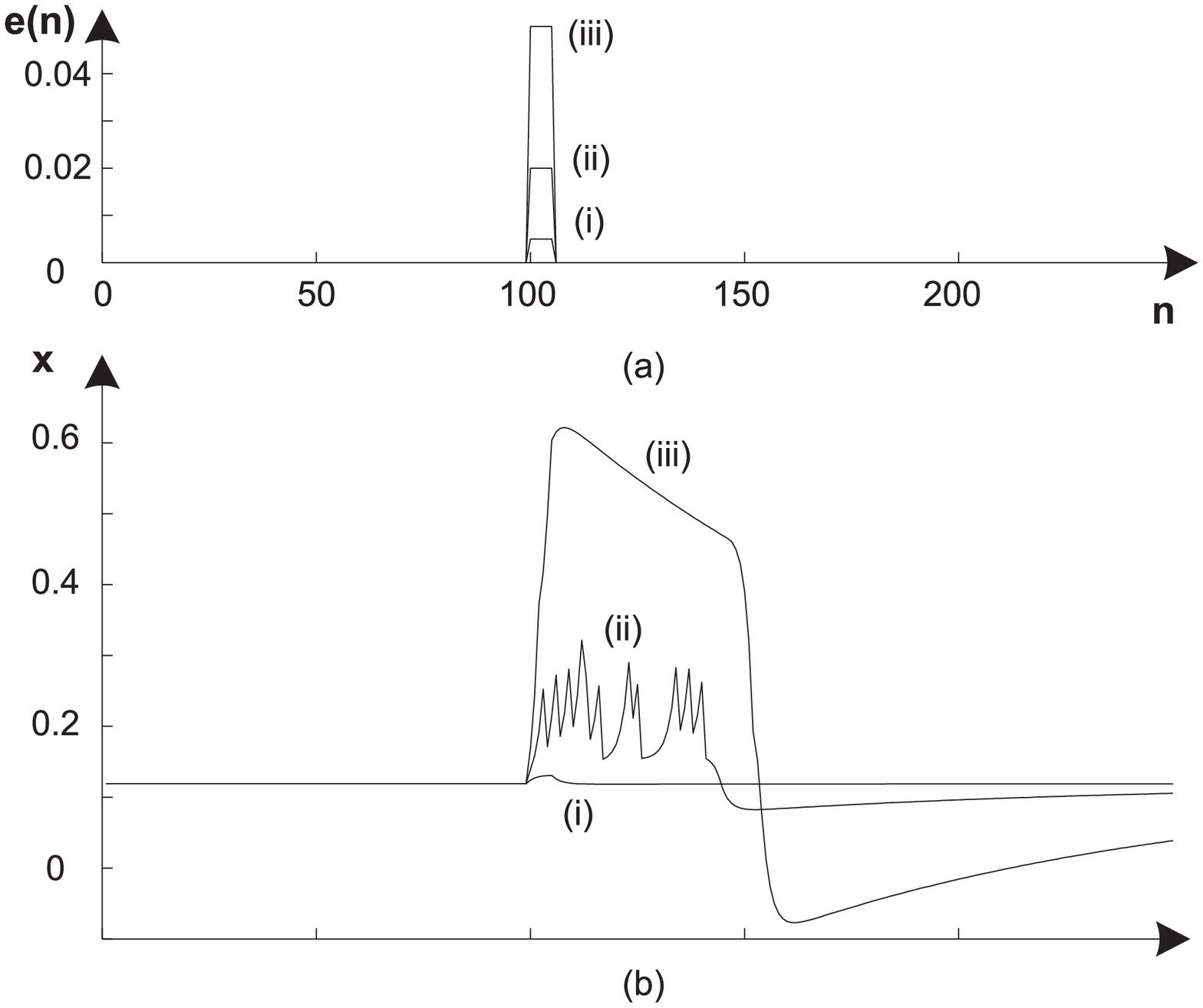} \\
   \includegraphics[width=0.7\textwidth]{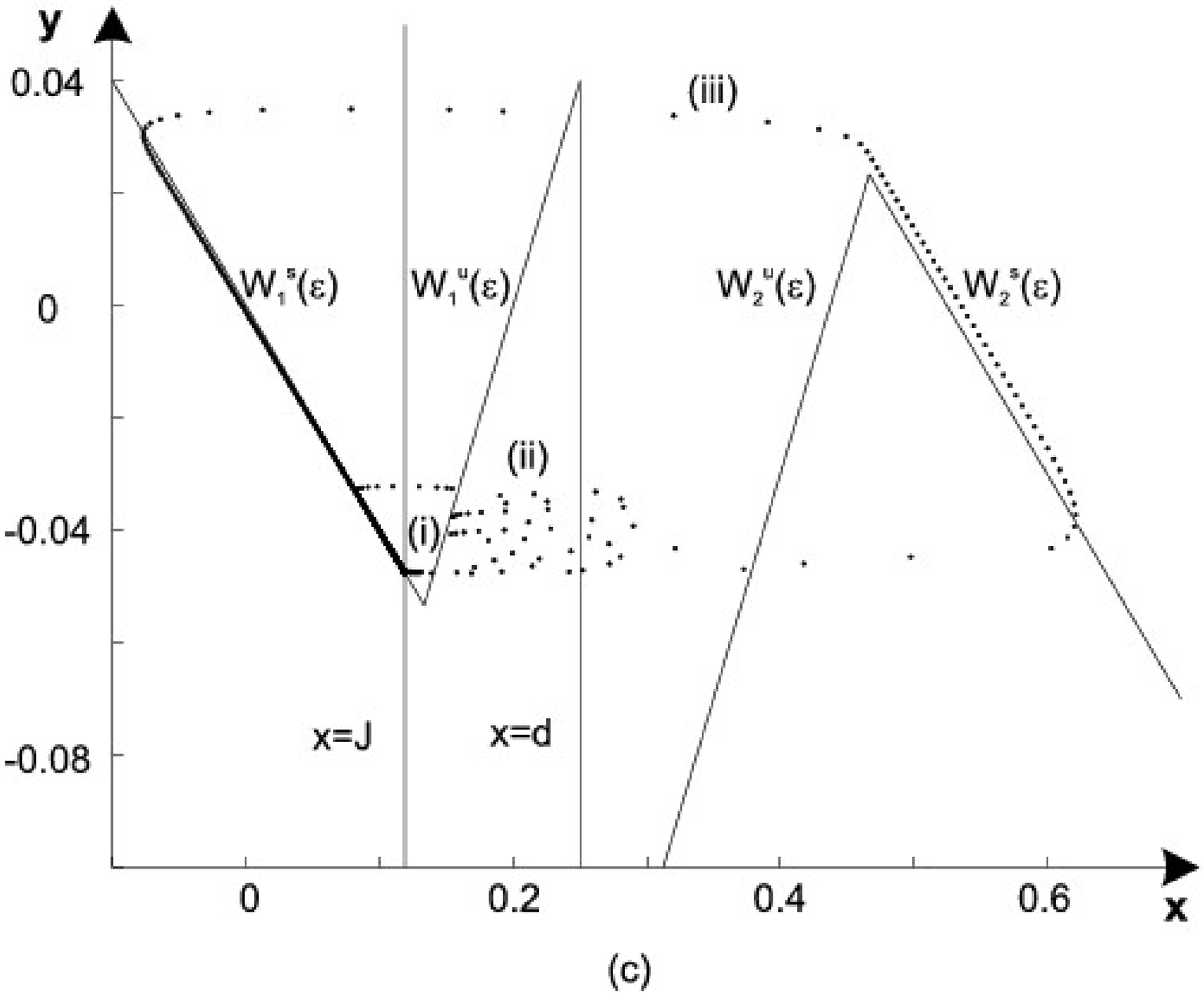} \\
   \caption{ Response of the model (\ref{eq:stpMp}) to positive
     pulse $e(n)$: (a) - three different amplitude of the
     stimulus; (b) - the behavior of variable $x$ (membrane
     potential of neuron); (c) - the phase plane.
     Parameter values: $J = 0.119, \varepsilon = 0.004,
     d = 0.25,\beta = 0.19, m_0 = 0.4, m_1 = 0.8,
     a = 0.2  $
     } \label{pic:PM_Ex2T}
\end{figure}

Studying response of the neurons to the influence
of external stimulus is one of important task of neuroscience
\cite{b:Kandel} associated with the problem of information
transmission in neural system. Usually external stimulus is
represented as the injection of electrical current into the
neuron. Let us suppose that the neuron is  not excited
initially, that is, it is in steady state (rest). In the model
(\ref{eq:stpMp}) such state of neuron corresponds to stable fixed
point $O$. Consider the response of the system (\ref{eq:stpMp}) to
pulse type stimulus. We  assume that the duration of each pulse is
small enough (see Fig. \ref{pic:PM_Ex2T}(a)) and its action is
equal to the instantaneous changing of the variable $x$ on the
pulse amplitude. Besides, we suppose here that $\varepsilon << 1$
and therefore the dynamics of the system (\ref{eq:stpMp}) is a
relaxation. For this parameter region the system (\ref{eq:stpMp})
has two thresholds. The first threshold is determined (see Fig.
\ref{pic:PM_Ex2T}(c)) by the thin layer of the slow motions near
the following invariant line
\begin{equation*}
     W_1^u(\varepsilon) = \left \{ (x,y):y=k_1x-b_1, \quad
         J_{min} < x <d  \right \}
\end{equation*}
where
\begin{equation*}
     b_1= m_1 a  - \frac{ \varepsilon J}{k_1}, \quad
         k_1=\frac{m_1}{2} + \sqrt{ \frac{m_1^2}{4}
         - \varepsilon }.
\end{equation*}
Analogously, the second threshold is defined (see Fig.
\ref{pic:PM_Ex2T}(c)) as the thin layer of the slow motions near
invariant line
\begin{equation*}
     W_2^u(\varepsilon) = \left \{ (x,y):y=k_1x-b_2, \quad
         d < x < J_{max}  \right \},
\end{equation*}
where
\begin{equation*}
     b_2= m_1 a  - \frac{ \varepsilon J}{k_1} + \beta.
\end{equation*}
Denote by the $x^e$ and $y^e$ ($y^e \approx F(J)$) the values of
the variables $x$ and $y$ after stimulation respectively. Let $E$
be the trajectory of the system (\ref{eq:stpMp}) with this initial
conditions. In other words $E$ is response of the system
(\ref{eq:stpMp}) to pulse input.
\par
(i) If the amplitude of  stimulus is not enough (Fig.
\ref{pic:PM_Ex2T}(a),(i)) for overcoming the first threshold, then the
maximum of the response will be about amplitude of the stimulus.
Therefore, in this case the generation of the actions  potential
does not take place.
\par
(ii) Let us increase the amplitude of the  stimulus as  it
breaks the first threshold but at the same time it is not enough
for overcoming of the second threshold (Fig.
\ref{pic:PM_Ex2T}(c),(ii)). In this case the fast motions of the
map $f$ will be close to the fast motions of the nap $g$ on interval $I$
for $y_0 \in Y_3 \bigcup Y_4$. And so, the trajectory of the map
$f$ perform some number of irregular oscillations around
discontinuity line $x=d$ (Fig. \ref{pic:PM_Ex2T}(c)(ii)). After
that, the trajectory $E$ within layer near $W_1^s(\varepsilon)$
tends to fixed point $O$ (Fig. \ref{pic:PM_Ex2T}(c)(ii)). Such
trajectory $E$ forms the region of phasic bursting activity
\cite{b:Izhikevich} with irregular number of spikes (Fig.
\ref{pic:PM_Ex2T}(b),(ii)).
\par
(iii) If the amplitude of the stimulus (Fig.
\ref{pic:PM_Ex2T}(a),(iii)) is enough for overcoming the second
threshold, then the point $x=x^e, y=y^e$ belongs to the region of
attractor of the invariant line $W_2^s(\varepsilon)$, where
\begin{equation*}
     W_2^s(\varepsilon) = \left \{ (x,y):y=-k_0 x-b_3, \quad
         x > J_{max}  \right \}
\end{equation*}
with
\begin{equation*}
     b_3= - m_0 a  - \frac{ \varepsilon J}{k_0} + \beta.
\end{equation*}
Therefore trajectory $E$ tends to thin layer of slow motions near
invariant line $W_2^s(\varepsilon)$. It moves within thin layer to
the neighborhood of the point $(x \approx J_{max}, y = F(J_{max})-
\beta)$ (Fig. \ref{pic:PM_Ex2T}(c),(iii)) and its motions continue
along fast motions. These motions are close to the trajectories of
the map $g$ for $y_0 \in Y_1$ (see Fig. \ref{pic:KL}(a)).
Therefore the trajectory $E$ tends to the layer near stable
invariant line $W_1^s(\varepsilon)$. After that the trajectory $E$
moves within thin layer near $W_2^S(\varepsilon)$ and it tends to
the fixed point $O$ (Fig. \ref{pic:PM_Ex2T}(c),(iii)). In this
case trajectory $E$ corresponds to phasic spike
\cite{b:Izhikevich} (Fig. \ref{pic:PM_Ex2T}(b),(iii)).

\subsection{ \label{sec:OtherOscill} Oscillatory modes of the neural
          activity}
\subsubsection{ \label{sec:CloseInvCurve} Close invariant curve
     and subthreshold oscillations }

\begin{figure}[t]
   \includegraphics[width=0.7\textwidth]{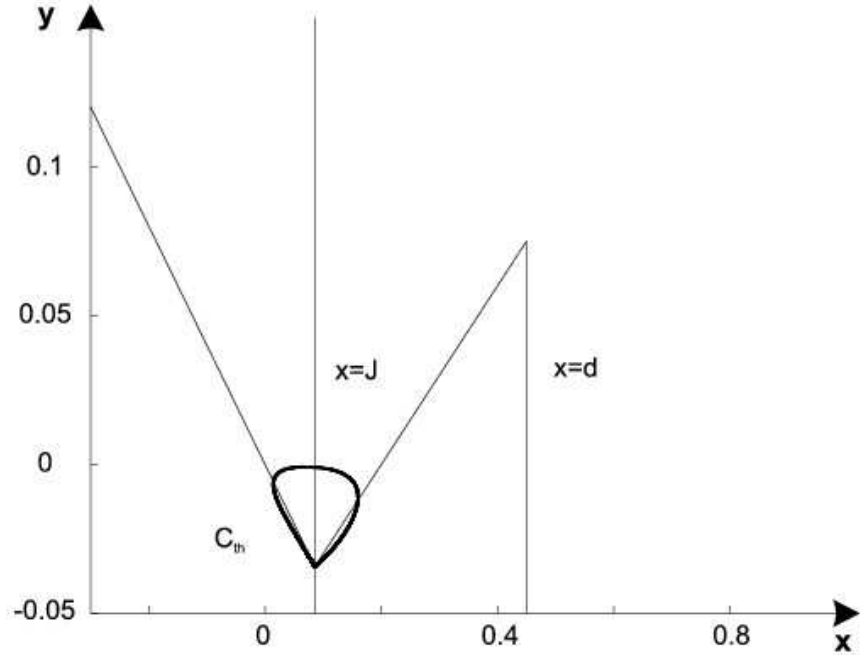} \\
   \includegraphics[width=0.7\textwidth]{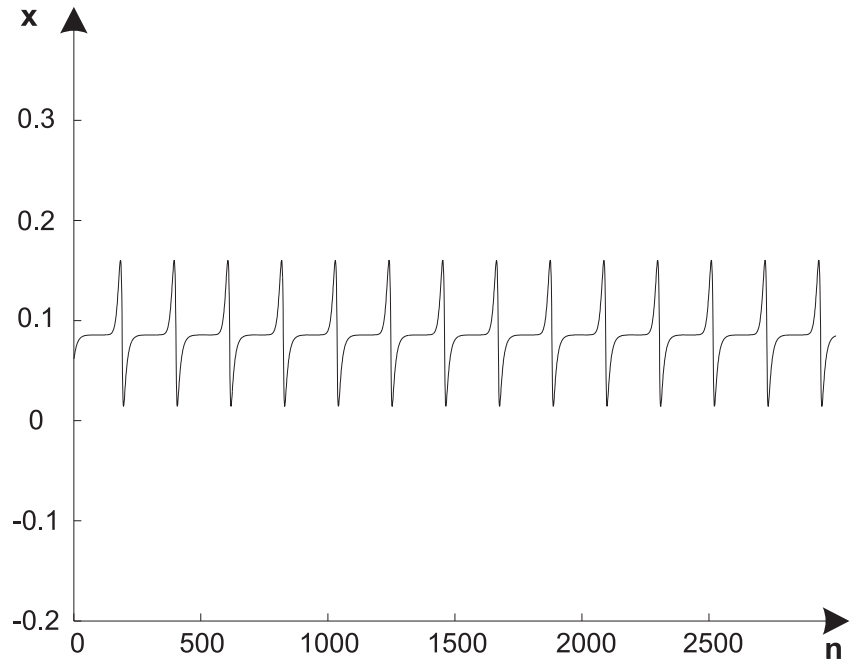} \\
   \caption{ (a) Close invariant curves on the phase plane $(x,y)$.
    (b) Subthreshold oscillations generated by map $f$.
     Parameter values: $d = 0.3, m_0 = 0.4, m_1 = 0.3,
       a = 0.2, J = 0.08572, \varepsilon = 0.025, \beta = 0.3$
       } \label{pic:PM_SO}
\end{figure}

Let us consider dynamics of the map $f$ under following conditions
on the parameters.
\begin{eqnarray*}
     \varepsilon < m_0, \quad m_0 > m_1^2 / 4 \\
     \varepsilon > \max \left \{ \frac{m_0^2}{4},
         \frac{m_1^2}{4} \right \} .
\end{eqnarray*}
One can see from the Jacobian matrix that in this case the fixed
point $O$ has a complex-conjugate multipliers. This point is
stable for $J < J_{min}$ and unstable for $J > J_{min}$. Therefore
the piece-wise map $f$ produces Neimark-Sacker like bifurcation
(in classical case of Neimark-Sacker bifurcation the map is
smooth). The fixed point $O$ is surrounded for $J>J_{min}$ by an
isolated stable attracting close curve $C_{th}$ (Fig.
\ref{pic:PM_SO}(a)). The oscillations corresponding to the
$C_{th}$ occur under the threshold of excitability of the neuron
and therefore it is called in neuroscience \cite{b:Llinas,
b:LlinasVortex}, subthreshold oscillations (Fig.
\ref{pic:PM_SO}(b)).

\subsubsection{ \label{sec:CloseInvCurve2} "Two-channel" chaotic
         attractor and chaotic spiking oscillations }

\begin{figure}[t]
   \includegraphics[width=0.7\textwidth]{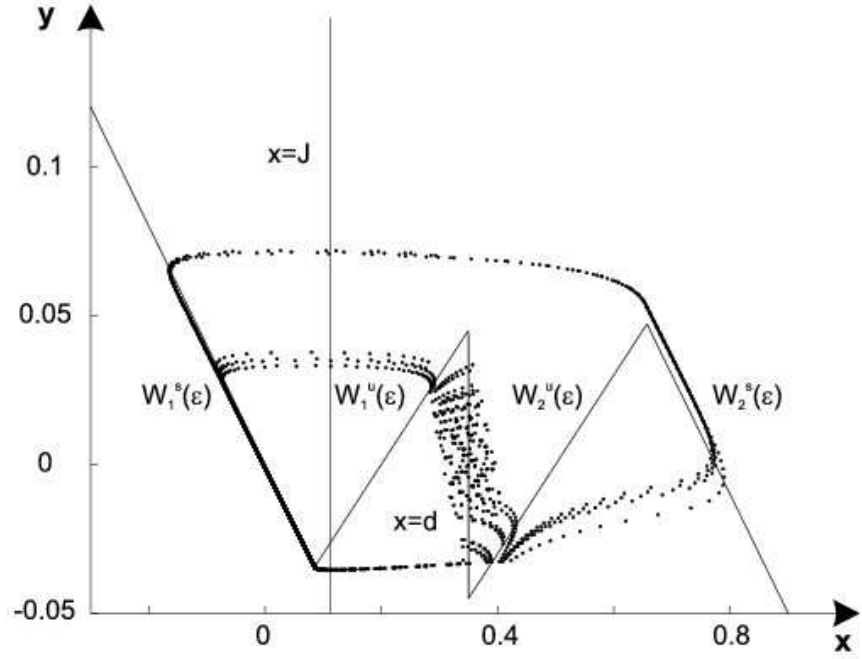} \\
   \includegraphics[width=0.7\textwidth]{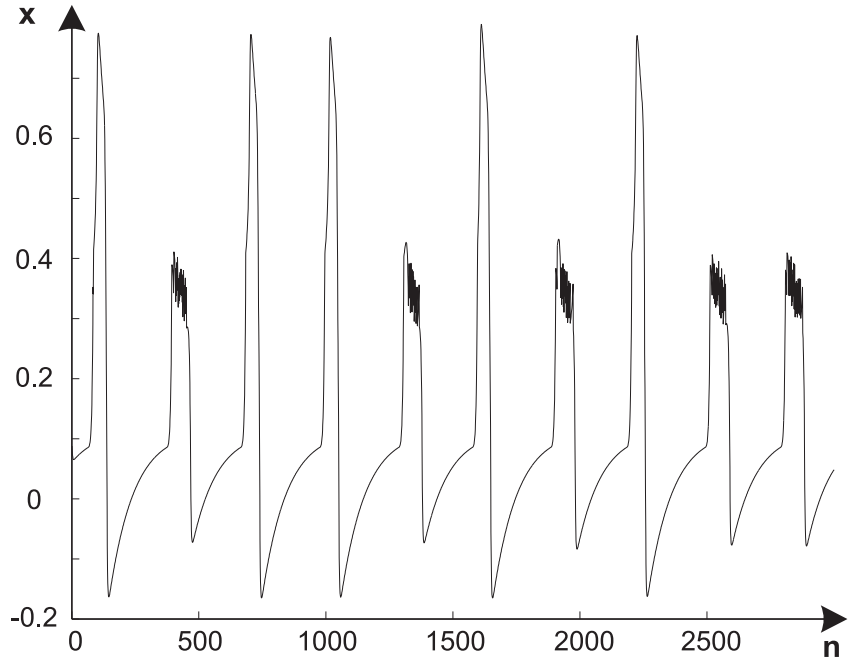} \\
   \caption{ (a) "Two-channels" chaotic attractor $A_{th}$. (b)
       Chaotic spiking against the background subthreshold
       oscillations. Parameter values: $J = 0.1123, \varepsilon =
       0.004, d = 0.3, m_0 = 0.4, m_1 = 0.3,
       a = 0.2,\beta = 0.09$} \label{pic:PM_CTS}
\end{figure}

Let us consider again the relaxation ($\varepsilon << 1$) dynamics
of the map $f$ in the case $J > J_{min}$, that is when fixed point
$O$ is unstable. Additionally we assume that the parameters of the
map $f$ has satisfied the following conditions
\begin{equation}\label{eq:CndForkMn}
     F(J_{min}) > F(d) - \beta,
\end{equation}

\begin{equation}\label{eq:CndForkMx}
     F(J_{max}) > F(d).
\end{equation}
In this case the invariant line $W_2^u(\varepsilon)$ separate the
fast motions on two flows (Fig. \ref{pic:PM_CTS}(a)) in the
neighborhood of the discontinuity line $x=d$. The first flow forms
the trajectory performing the chaotic oscillations near the line
$x=d$ (Fig. \ref{pic:PM_CTS}(a)). Their dynamics are close to the
dynamics of the map $g$ on interval $I$ for $y_0 \in Y_3 \bigcup
Y_4$. The second flow consists of the trajectories overcoming the
second threshold (Fig. \ref{pic:PM_CTS}(a)). It moves to
neighborhood of the stable invariant line $W_2^s(\varepsilon)$.
After that these trajectory tends to the stable invariant line
$W_1^s(\varepsilon)$ and the described process is repeated. These
trajectories form chaotically switching flow from one to other. As
a result is the appearance of a chaotic attractor $A_{th}$ on
phase plane (Fig. \ref{pic:PM_CTS}(a)). The fractal dimension
$d_f(A_{th})=1.30335$. The attractor $A_{th}$ determines chaotic
regime of spiking activity over the background of the subthreshold
oscillations (Fig. \ref{pic:PM_CTS}(b)).

\subsubsection{ \label{sec:CloseInvCurve3} Close invariant curve
                 and tonic spiking}
\begin{figure}[t]
   \includegraphics[width=0.7\textwidth]{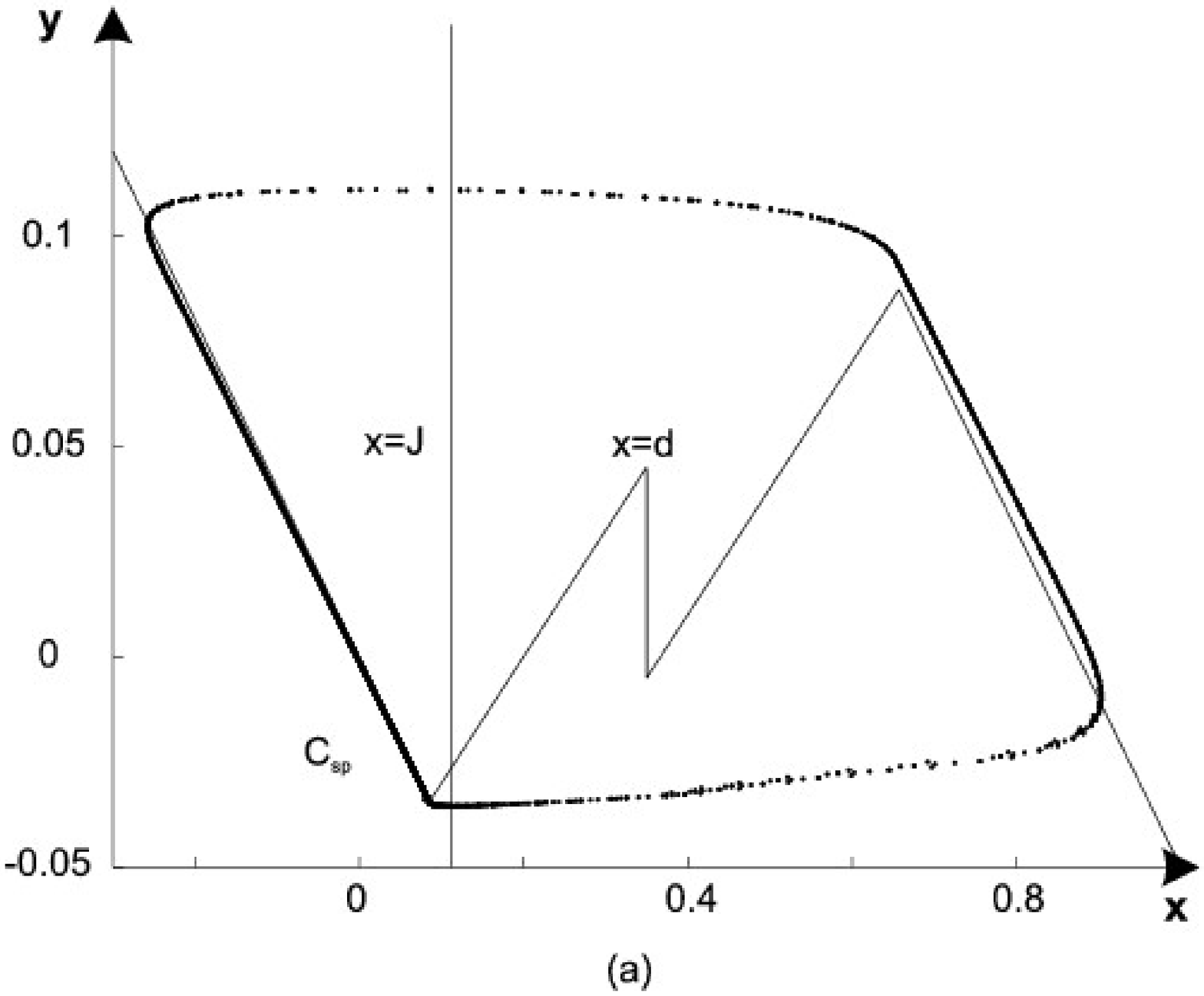} \\
   \includegraphics[width=0.7\textwidth]{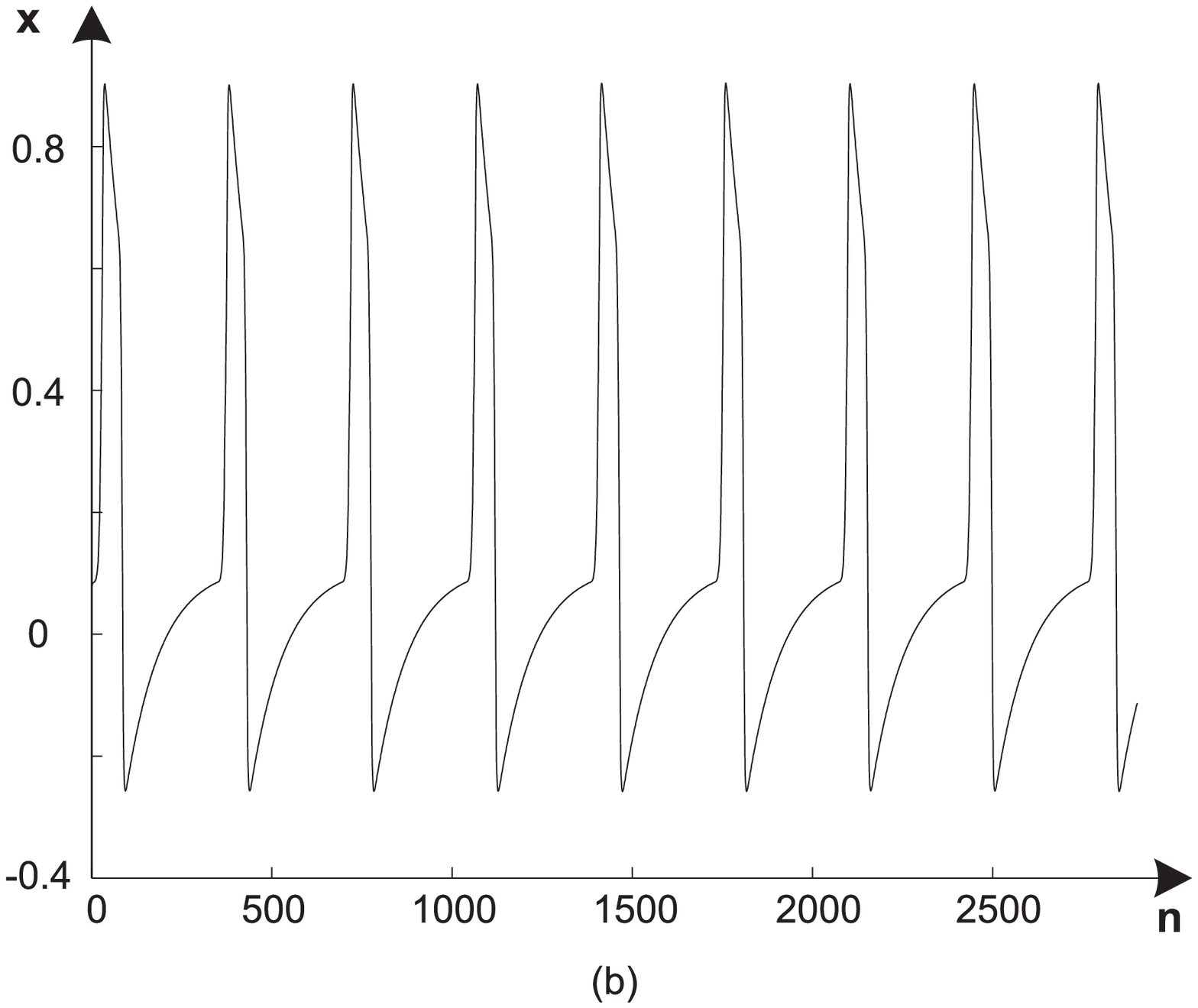} \\
   \caption{ (a) Close invariant curve. (b) Tonic spiking.
     Parameter values: $J = 0.1123, \varepsilon = 0.004
       d = 0.3, m_0 = 0.4, m_1 = 0.3,
       a = 0.2,\beta = 0.05$ } \label{pic:PM_TS}
\end{figure}

Let  the parameters of the map $f$  satisfy the same conditions as
in the case of previous subsection \ref{sec:CloseInvCurve} with
exception of inequality (\ref{eq:CndForkMn}). In this case the
parameter $\beta$ is small enough. Therefore the trajectories with
initial conditions from neighborhood of $W_{1,2}^s(\varepsilon)$
do not change direction of motion when they intersect the
discontinuity line $x=d$. This leads to the appearance on the phase
plane of the motions between layer near $W_1^s(\varepsilon)$ and
$W_2^s(\varepsilon)$. Such dynamics leads to forming close
invariant curve $C_{sp}$ (Fig. \ref{pic:PM_TS}(a)). So there exists
only one attractor on the phase plane formed by this invariant
closed curve. This determines tonic spiking regimes of neural
activity (Fig. \ref{pic:PM_TS}(b)).

\section{ \label{sec:Conclusion} Conclusion \protect\\}
A new phenomenological model of neural activity is proposed. The
model can reproduce basic activity modes such as spiking, chaotic
spiking-bursting, subthreshold oscillations etc. of real
biological neurons. The model is a discontinous two-dimensional
map based  on the discrete version of the FitzHugh -
Nagumo system and dynamical properties the Lorenz-like map. We
have shown that the dynamics of our model display both regular
and chaotic behavior. We have studied the underlying mechanism of
the generation of chaotic spiking-bursting oscillations.
Sufficient condition  for existence chaotic attractors in the
phase plane are obtained. In spite of idealization, the dynamical
modes which are demonstrated in our model are in agreement with
the neural activity regimes experimentally found in real
biological systems. For example, subthreshold oscillations (see
Fig. \ref{pic:PM_SO}(b)) is a basic regime of inferior olive (I.O.)
neurons \cite{b:Llinas}. Inferior olive neurons belong to the
olivo-cerebellar network which plays a key role
\cite{b:LlinasVortex} in organization of vertebrate motor control.
It is also typical for I.O. neuron  \cite{b:Bernardo} a spiking
regime over the chaotic subthreshold oscillations (see Fig.
\ref{pic:PM_CTS}).  The spiking-bursting activity is significant
for many types of neurons,  in particular in hippocampal
pyramidal cell \cite{b:Wang} and thalamic cells \cite{b:Deschenes}.
\par
The table summarizes results on gallery of behavior of neural
activity showed by our model. We hope that our model will be
useful to understand the mechanism of neural pattern formation in
large networks.
\begin{center}
     \begin{tabular}{|c|c|}
     \multicolumn{2}{c} { \bf TABLE } \\
     \hline
     { \bf Parameters }& { \bf  The regimes of neuronal activity }\\
     \hline &\\
     \begin{minipage}{0.35\textwidth}
         \begin{center}
         $J < J_{min}$, $\varepsilon << 1$
         \par
         $F(J)  < F(J_{max})-\beta$,
         \par
         $F(J_{max})-\beta > F(d)$ (spike)
         \par
         $F(J) > F(d) -\beta$ (bursts)
         \end{center}
     \end{minipage}
                         &   \begin{minipage}{0.3\textwidth}
                                 \begin{center}
                                     Phasic spikes and bursts \par
                                     \includegraphics[width=1\textwidth]
                                         {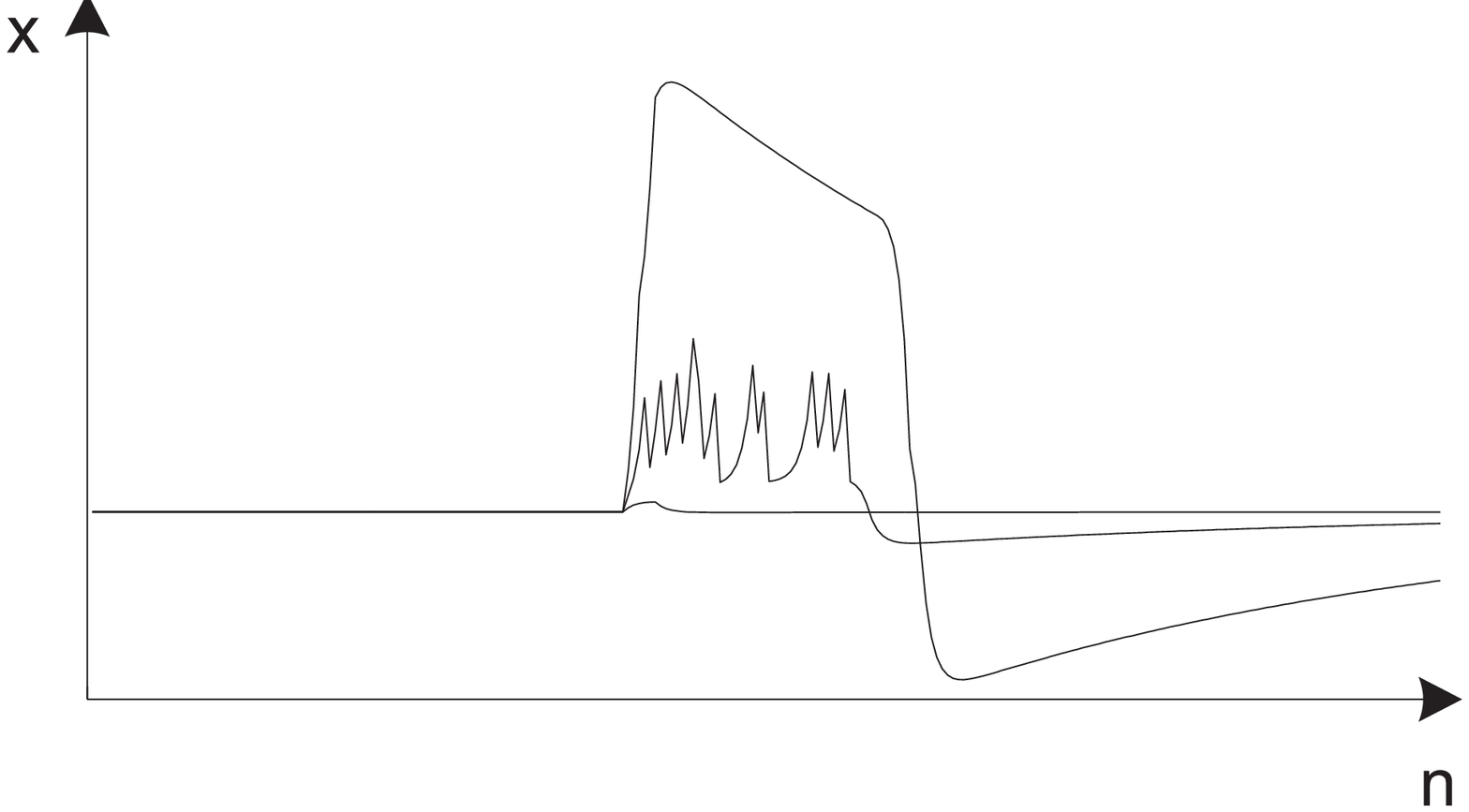}
                                 \end{center}
                             \end{minipage}\\
     \hline &\\
     \begin{minipage}{0.35\textwidth}
         \begin{center}
             $J > J_{min}$, $\varepsilon < m_0 $ \par
             $m_0 > m_1^2 / 4$ \par
             $\varepsilon > \max \left \{ \frac{m_0^2}{4},
                     \frac{m_1^2}{4} \right \}$
         \end{center}
     \end{minipage}
                         &   \begin{minipage}{0.3\textwidth}
                                 \begin{center}
                                     Subthreshold oscillations \par
                                     \includegraphics[width=1\textwidth]
                                         {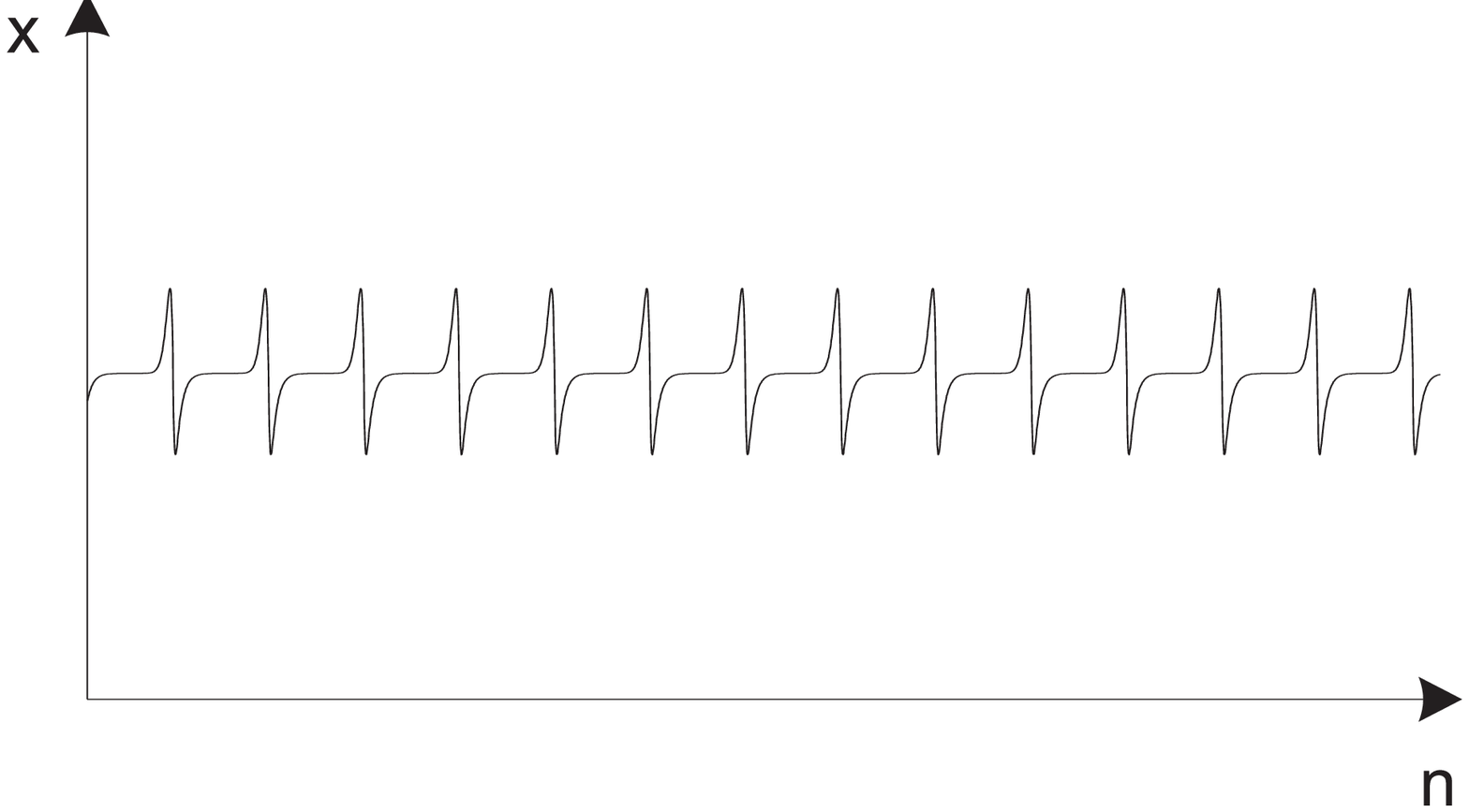}
                                 \end{center}
                             \end{minipage}\\
     \hline &\\
     \begin{minipage}{0.35\textwidth}
         \begin{center}
             Inequalities (\ref{eq:CndInvEpsJ}).
         \end{center}
     \end{minipage}
                         &   \begin{minipage}{0.3\textwidth}
                                 \begin{center}
                                     Chaotic bursting oscillations \par
                                     \includegraphics[width=1\textwidth]
                                         {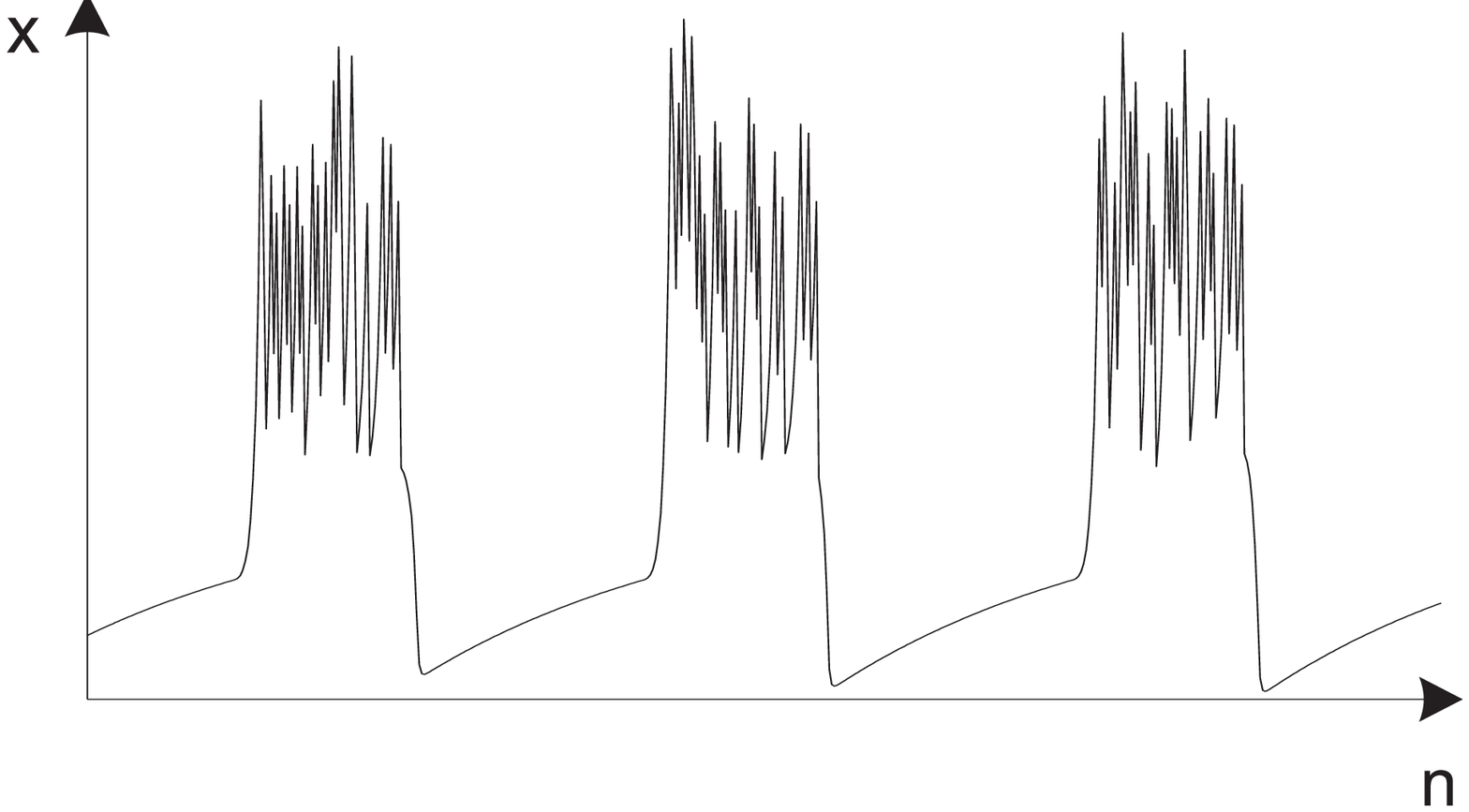}
                                 \end{center}
                             \end{minipage}\\
     \hline &\\
     \begin{minipage}{0.35\textwidth}
         \begin{center}
             $J > J_{min}$, $\varepsilon << 1$ \par
             $F(J_{min}) > F(d) - \beta$ \par
             $F(J_{max}) > F(d)$
         \end{center}
     \end{minipage}
                         &   \begin{minipage}{0.3\textwidth}
                                 \begin{center}
                                     Chaotic spiking \par
                                     \includegraphics[width=1\textwidth]
                                         {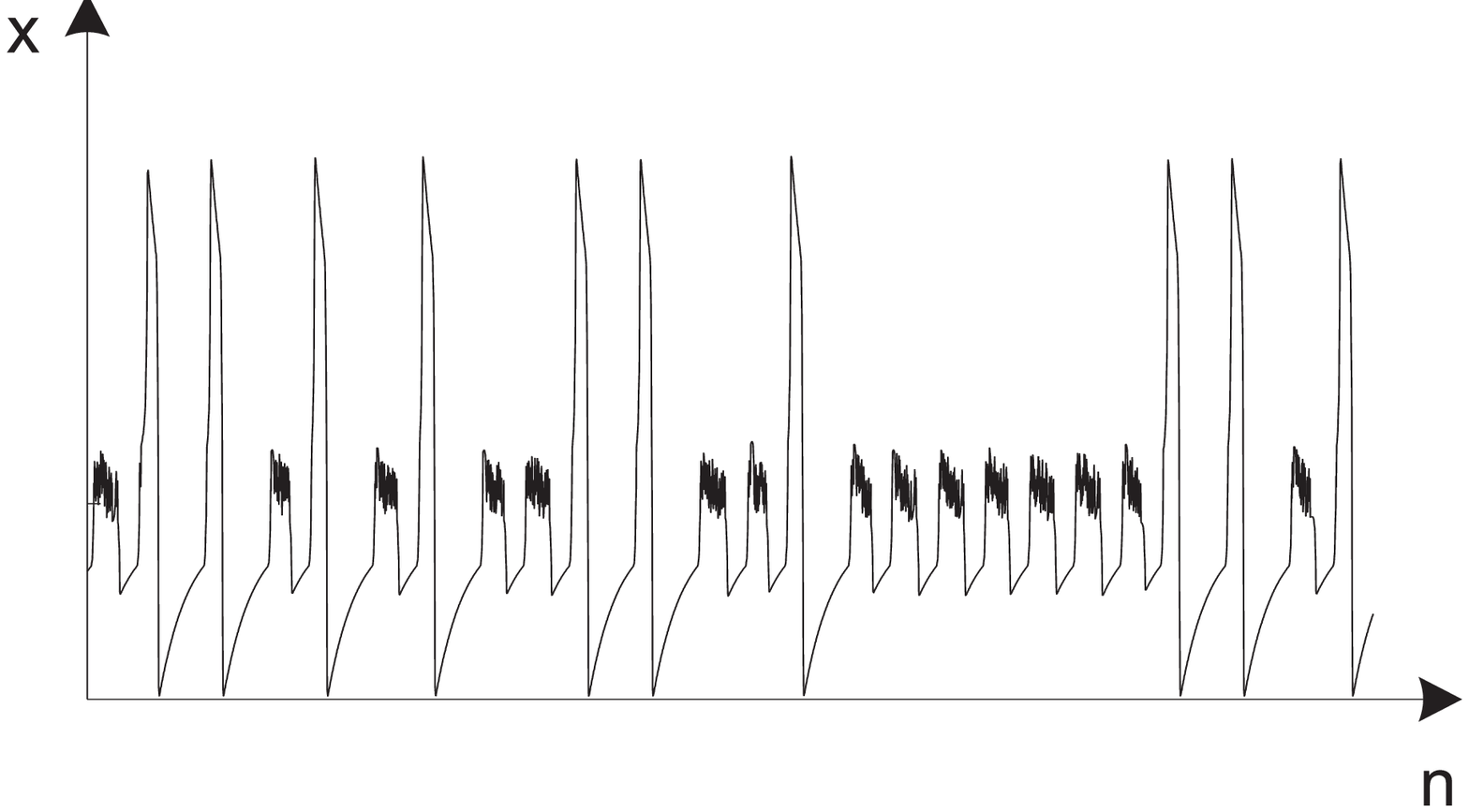}
                                 \end {center}
                             \end{minipage}\\
     \hline &\\
     \begin{minipage}{0.35\textwidth}
         \begin{center}
             $J > J_{min}$, $\varepsilon << 1$ \par
             $F(J_{min}) < F(d) - \beta$.
         \end{center}
     \end{minipage}
                         &   \begin{minipage}{0.3\textwidth}
                                 \begin{center}
                                     Tonic spiking \par
                                     \includegraphics[width=1\textwidth]
                                         {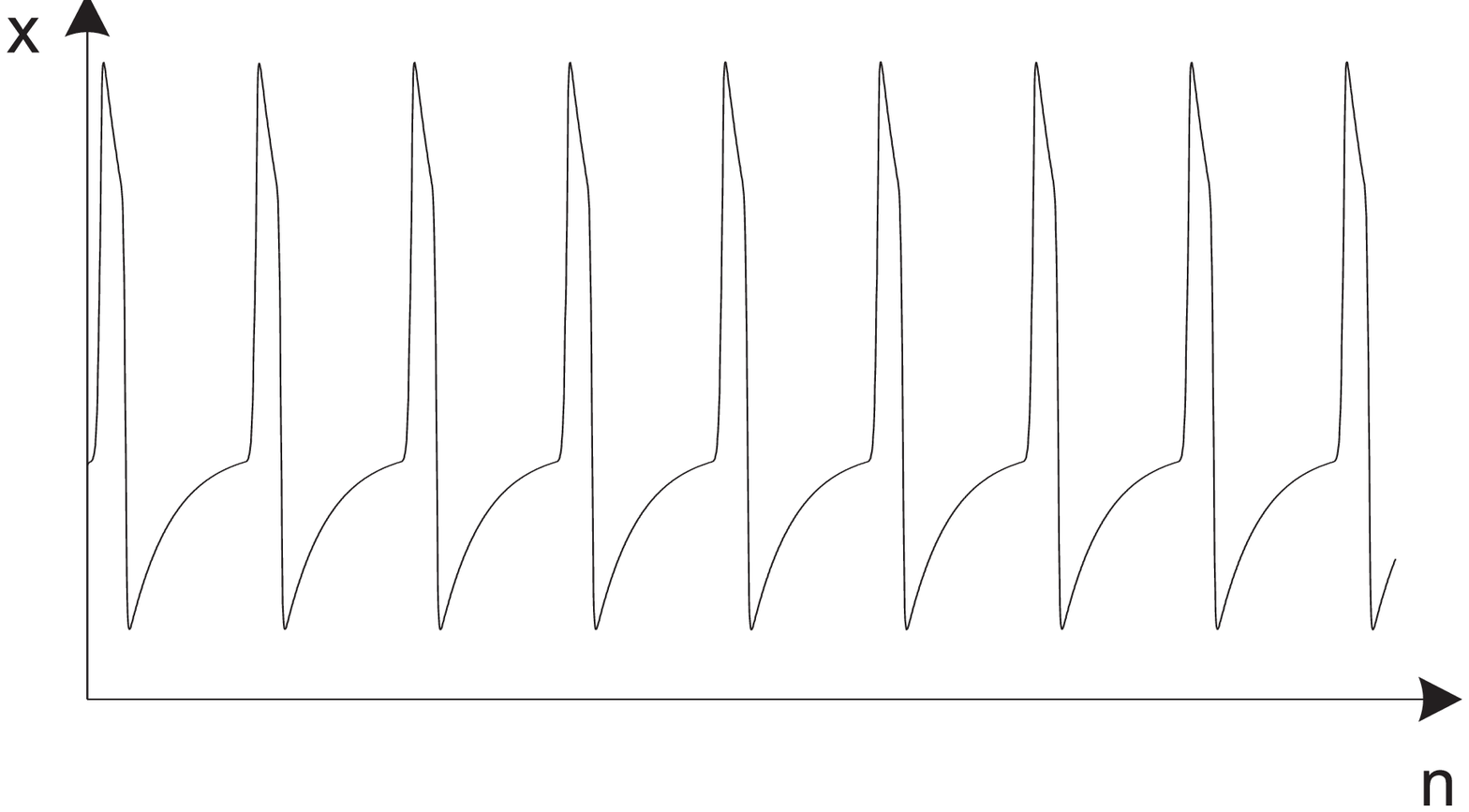}
                                 \end{center}
                             \end{minipage}\\
     \hline
     \end{tabular}
\end{center}
\section*{ Acknowledgments}
This work was supported partly by University Paris 7-Denis Diderot
and in part by the Russian Foundation for Basic Research (grant
06-02-16137) and Leading Scientific Schools of the Russian
Federation (Scientific School 0 7309.2006.2).

\appendix
\section*{Appendix. The equations of the boundaries of the invariant region}
The boundary $\Gamma$  is given by
\begin{eqnarray*}
     &\Gamma_1 =& \left\{ (x,y): d \leq x \leq J_{max}, \quad
         y = - \frac{ \sqrt{\varepsilon}}{m_0} (J - J_{min})
         - m_0 J - k_1 (d-J)
         \right\} \\
     &\Gamma_2 =& \left\{ (x,y): J \leq x \leq d, \quad
         y = -p(x-J) - \frac{ \sqrt{\varepsilon}}{m_0}(J - J_{min})
         - m_0 J \right\} \\
     &\Gamma_3 =& \left\{ (x,y): J_{min} \leq x \leq J , \quad
         y = - \frac{ \sqrt{\varepsilon}}{m_0}( x - J_{min})
         - m_0 J \right\} \\
     &\Gamma_4 =& \left\{ (x,y): -J_1 \leq x \leq J_{min}, \quad
          y = -k_0 (x - J_{min}) - m_0 J \right\} \\
     &\Gamma_5 =& \left\{ (x,y): -J_1 \leq x \leq - J_0, \quad
         \sqrt{ \varepsilon } y = (x+J_0)
         + \sqrt{\varepsilon} m_0 J_0 \right\} \\
     &\Gamma_6 =& \left\{ (x,y): -J_0 \leq x \leq J_{min}, \quad
         y = m_0 J_0 \right\} \\
     &\Gamma_7 =& \left\{ (x,y): J_{min} \leq x \leq J, \quad
         y = -2 \sqrt{ \varepsilon } ( x - d)
         + F(d) + \sqrt{\varepsilon}(d-J) \right\} \\
     &\Gamma_8 =& \left\{ (x,y): x = J_{max},
          y_1 \leq y \leq y_2
           \right\}
\end{eqnarray*}
with
\begin{eqnarray*}
     y_1 &=& - \frac{ \sqrt{\varepsilon}}{m_0} (J - J_{min})
         - m_0 J - k_1 (d-J) \\
     y_2 &=& -2 \sqrt{ \varepsilon } ( J_{max} -  d)
         + F(d) + \sqrt{ \varepsilon } (d-J)
\end{eqnarray*}
\begin{equation*}
     p = \begin{cases}
         \frac{B}{2(d-J)}
             - \sqrt{ \frac{B^2}{4(d-J)^2} -\varepsilon },
             \quad &\text{if} \quad
             \sqrt{\varepsilon} \leq \frac{B}{2(d-J)} \\
         \frac{B-m_1(J_{max}-d)}{d-J},
             \quad &\text{if} \quad
             \sqrt{\varepsilon} > \frac{B}{2(d-J)}
     \end{cases},
\end{equation*}
\begin{equation*}
     B = \beta - F(d) - m_0 J,
\end{equation*}
\begin{equation*}
     J_0 = \frac{ \sqrt{\varepsilon} \left[ (d - J)
         + 2(d-J_{min})\right]
         +F(d) }{m_0},
\end{equation*}
\begin{equation*}
     J_1 = \frac{ (1+\sqrt{\varepsilon}m_0) J_0
         - \sqrt{\varepsilon} (k_0 J_{min} - m_0 J ) }
         {1+k_0\sqrt{\varepsilon}}
\end{equation*}
The boundary $\gamma$ is given by
\begin{eqnarray*}
     &\gamma_1=& \left\{ (x,y) : J_{min} \leq x \leq J,
         \quad y = -m_0 J_{min} \right\} \\
     &\gamma_2=& \left\{ (x,y) :  x = J_{min},
         \quad -m_0 J_{min} \leq y \leq m_1(J-a) \right\} \\
     &\gamma_3=& \left\{ (x,y) :  J_{min} \leq x \leq J,
         \quad y = m_1(J-a) + \sqrt{\varepsilon}
         (x-J_{min}) \right\} \\
     &\gamma_4=& \left\{ (x,y) : J \leq x \leq J +
         \frac {\sqrt{ \varepsilon } (J - J_{min})}{m_1},
         \quad y = m_1(J-a)
         + \sqrt{\varepsilon}(J-J_{min}) \right\} \\
     &\gamma_5=& \left\{ (x,y) :  x = J +
         \frac {\sqrt{ \varepsilon } (J - J_{min})}{m_1},
         \quad y_3 \leq y \leq y_4 \right\} \\
     &\gamma_6=& \left\{ (x,y) : J \leq x \leq J +
         \frac {\sqrt{ \varepsilon } (J - J_{min})}{m_1},
         \quad y = \sqrt{\varepsilon} (x-J)
         -m_0 J_{min} \right\}
\end{eqnarray*}
with
\begin{eqnarray*}
     y_3 &=& \frac{ \varepsilon(J-J_{min})}{m_1}
         - m_0 J_{min} \\
     y_4 &=& m_1(J-a) + \sqrt{\varepsilon}(J-J_{min})
\end{eqnarray*}

\newpage

\begin{figure}[t]
   \includegraphics[width=0.75\textwidth]{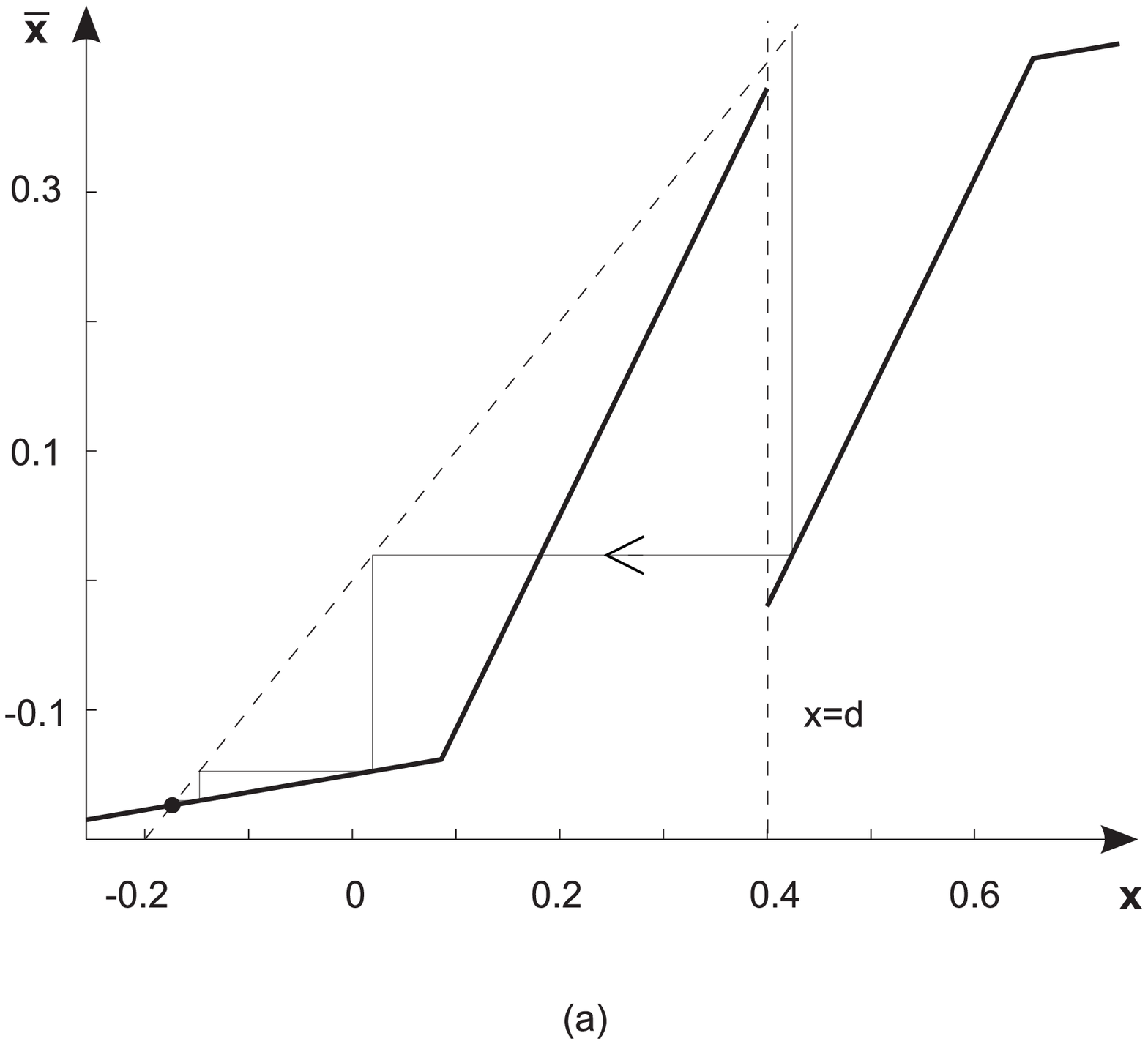}
   \includegraphics[width=0.75\textwidth]{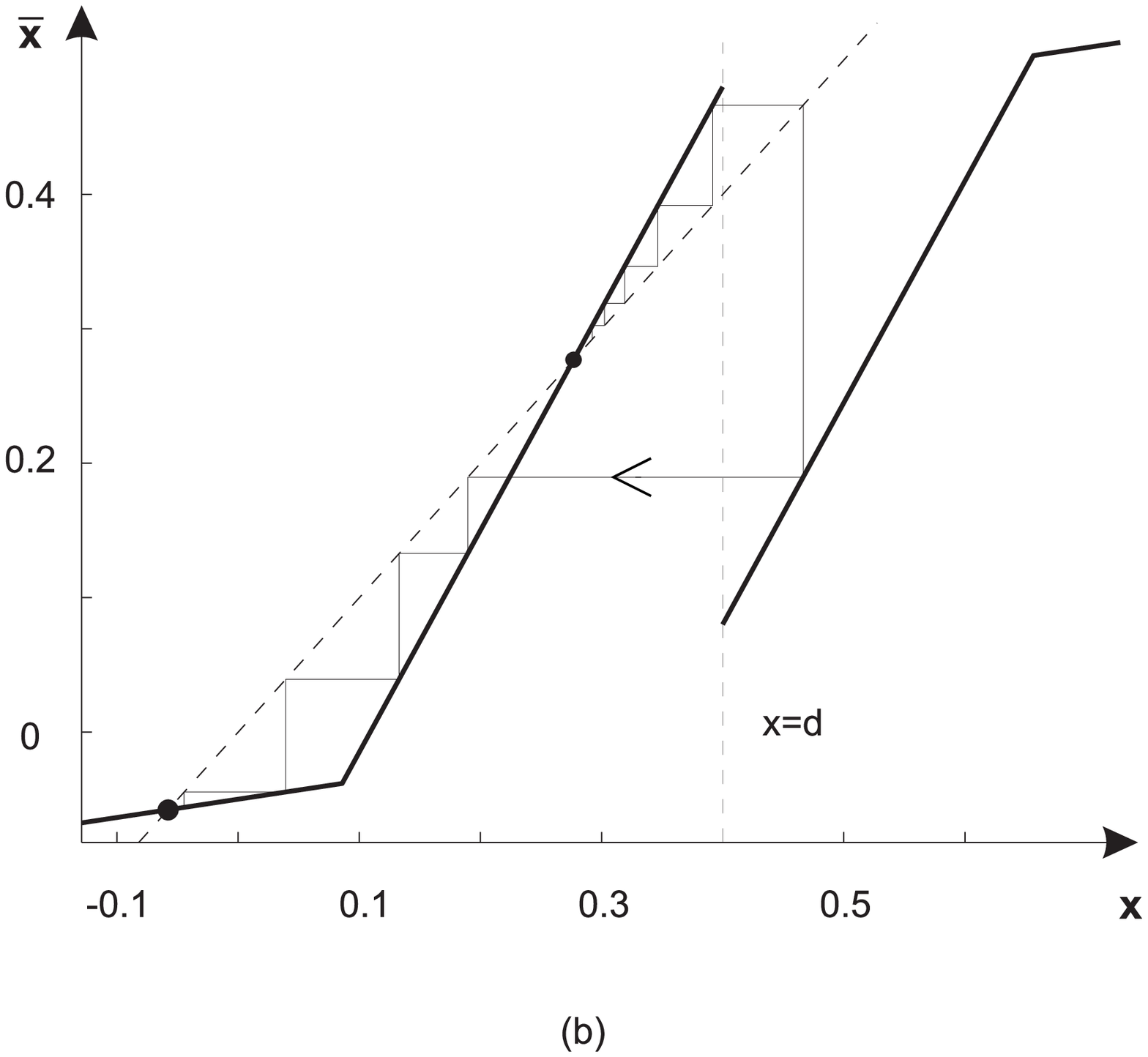} \\
\end{figure}

\begin{figure}[t]
   \includegraphics[width=0.75\textwidth]{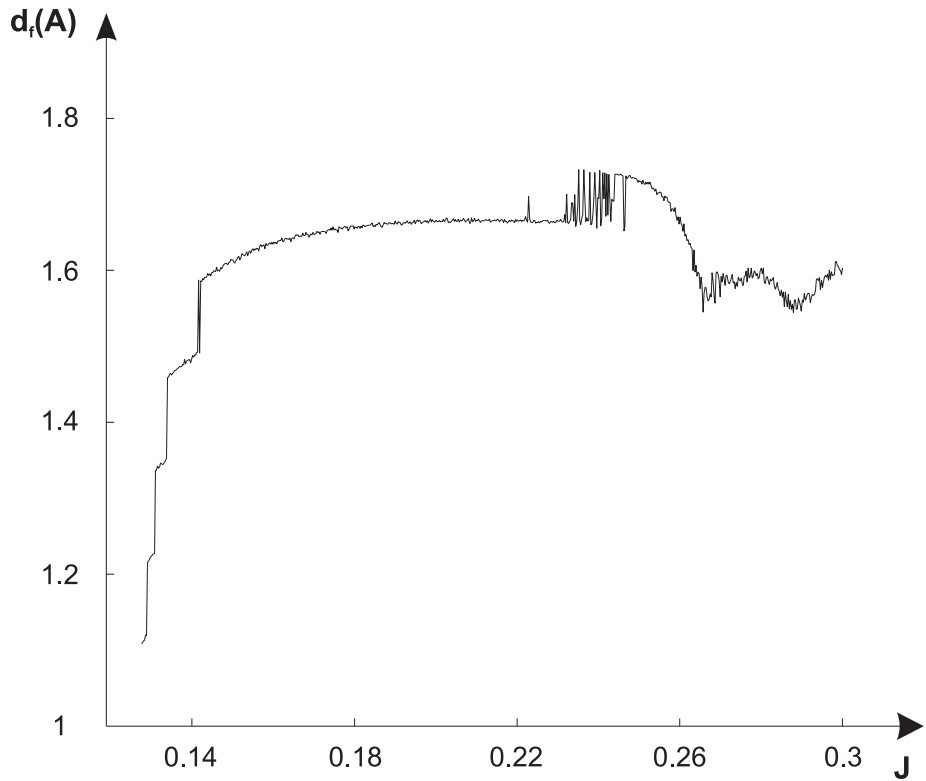} \\
   \caption{ Fractal dimension $d_f(A)$ of the attractor $A$ versus
     parameter $J$. Parameter values: $m_0 = 0.4,
     m_1 = 0.65, a = 0.2, d = 0.3,
     \beta = 0.25,  \varepsilon = 0.002 $
     }\label{pic:ChAtt2}
\end{figure}

\begin{figure}[t]
   \includegraphics[width=0.75\textwidth]{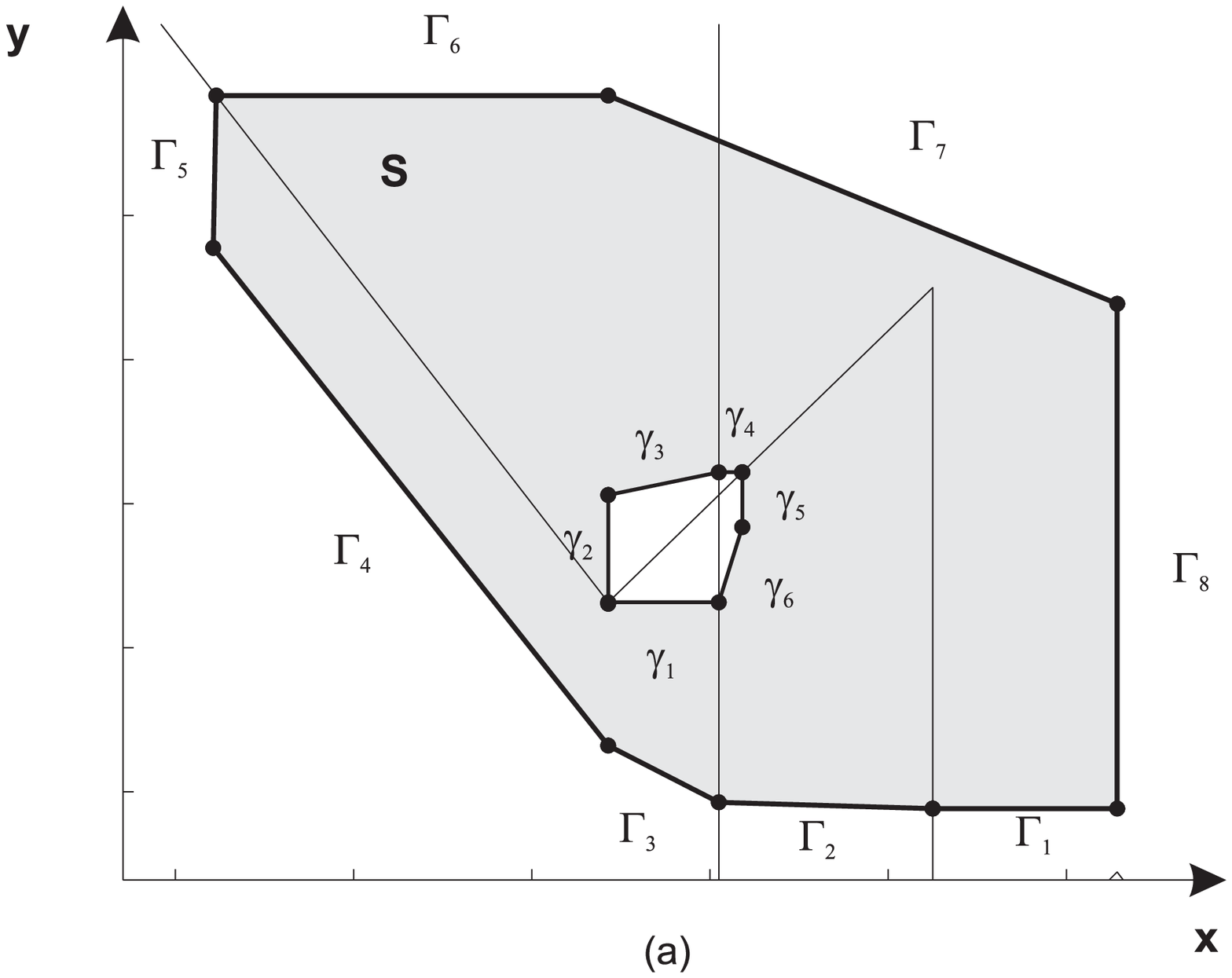} \\
   \includegraphics[width=0.75\textwidth]{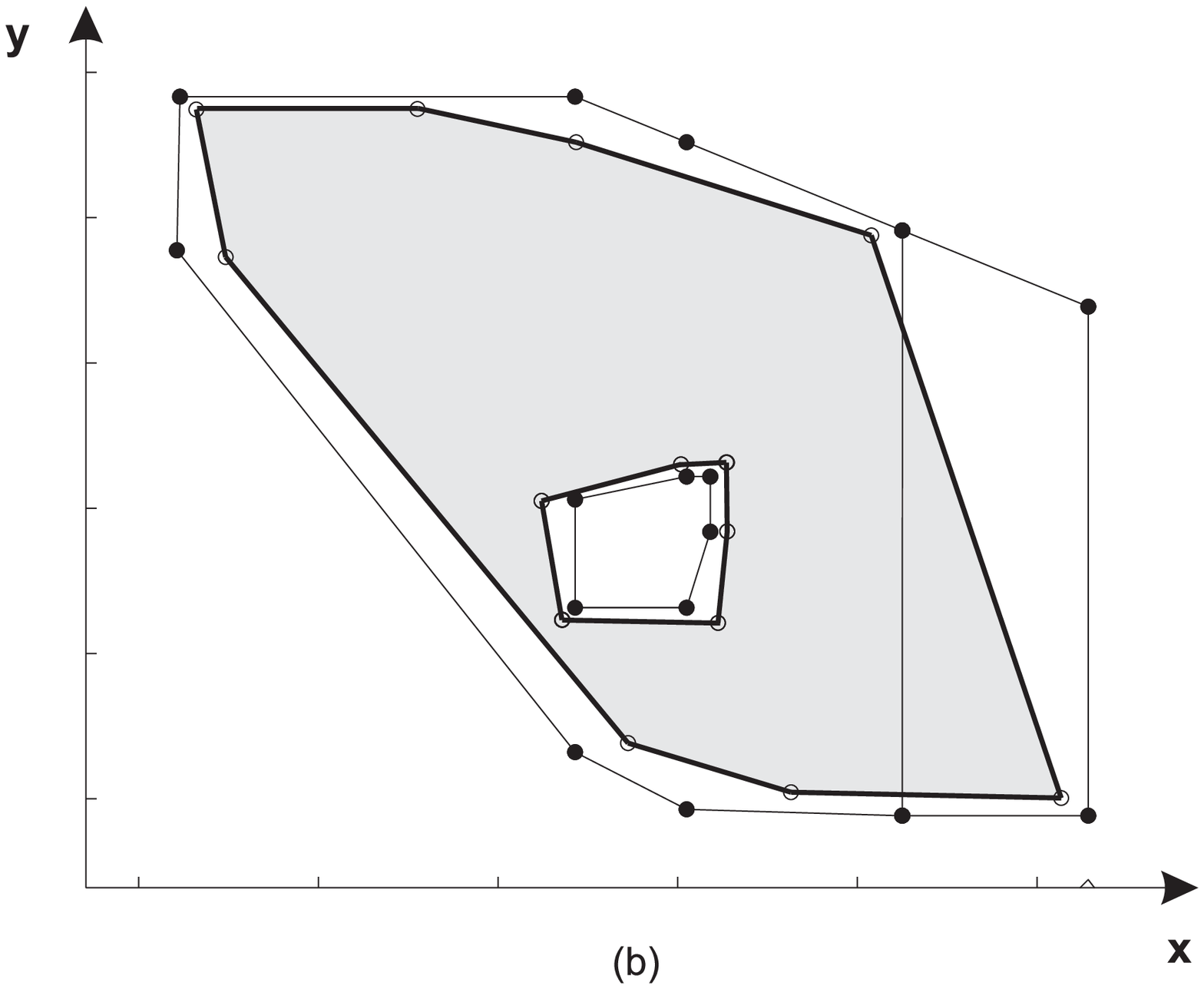}\\
\end{figure}

\begin{figure}[t]
   \includegraphics[width=0.75\textwidth]{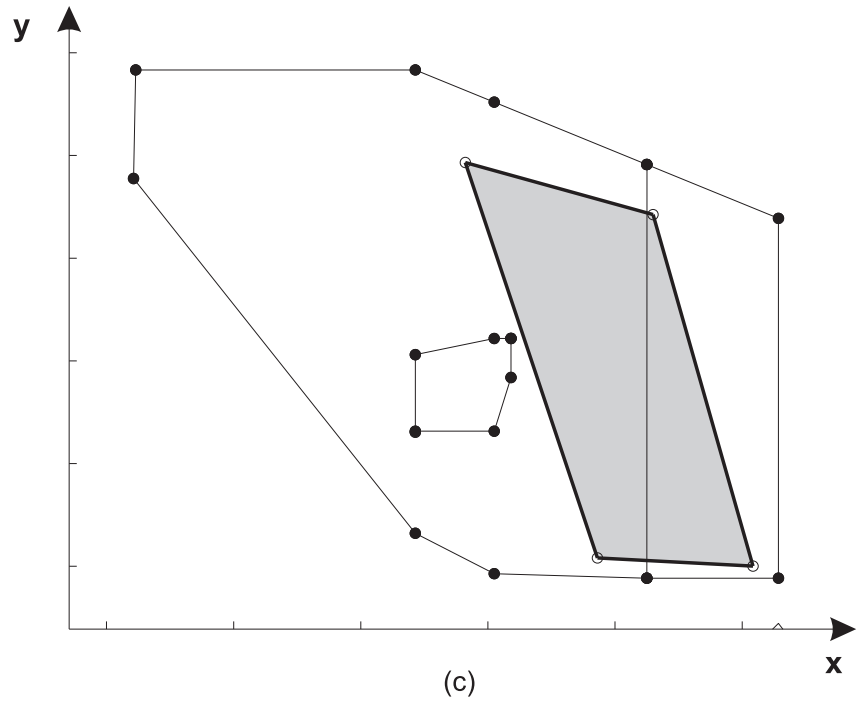}\\
   \caption{ (a) The ring-like invariant region $S$.
     (b) The image of the region $S \bigcap \{x \leq d \}$
     under action of the map $f_1$; (c) The image of the
     region $S \bigcap \{x \geq d\}$ under action of the
     the map $f_2$.
     } \label{pic:PM_InvRg}
\end{figure}

\begin{thebibliography}{}
     \bibitem{b:Kandel}
         \textit{ E.R. Kandel, J.H. Schwartz, T.M Jessell, }
         Principles of neural science, Prentice-Hall Int. Inc.,
         1991.
     \bibitem{b:Rabinovich}
         \textit{ M.I. Rabinovich, P. Varona, A.I. Selverston,
             H.D.I. Abarbanel, }
         Dynamical principles in neuroscience,
         Reviews of Modern Physics, 78(4) (2006), 1213.
     \bibitem{b:IandFModel}
         \textit{ H.R. Wilson, J.D.Cowan }
         Excitatory and inhibitory interaction in localized
         population of model neurons. Biophys. J., 12 (1972), pp
         1-24.
     \bibitem{b:FitzHugh} \textit{R. FitzHugh} Mathematical models of the
         threshold phenomena in the nerve membrane. Bull Math.
         Biohys. v.17, pp. 257-287 (1955)
     \bibitem{b:HingmarshRose}
         \textit{J.L. Hindmarsh , R.M. Rose },
         A model of neuronal bursting using three coupled first order
         differential equations: Philos. Trans R. Soc. London, Ser.
         B221, 87-102 (1984).
     \bibitem{b:MorrisLekar}
          \textit{C. Morris, H. Lecar},
         Voltage oscillations in the barnacle giant muscle fiber.
         Biophys. J. v25, 1981, p. 87
     \bibitem{b:Chialvo}
         \textit{ D.R. Chialvo, }
         Generic excitable dynamics on the two-dimensional map,
         Chaos Solitons Fract. 5 (1995) 461-480.
     \bibitem{b:Kinouchi}
         \textit{ O. Kinouchi, M. Tragtenberg, }
         Modeling neurons by simple maps, Int. J. Bifurcaton Chaos
         6, N 12 a (1996), 2343-2360.
     \bibitem{b:Kuva}
         \textit{S.Kuva, G.Lima, O.Kinouchi,
             M.Tragtenberg, A. Roque, }
         A mimimal model for excitable and bursting element. Neurocomputing
         38-40, pp. 255-261 (2001).
     \bibitem{b:DeVries}
         \textit{ G. de Vries, }
         Bursting as an emergent phenomenon in coupled chaotic
         maps, Phys. Rev. E 64 (2001) 051914.
     \bibitem{b:RulkovTim}
         \textit{N.F. Rulkov, I.Timofeev, M. Bazhenov,}
         Oscillations in large-scale cortical networks:
         map-based model. J. of Computational Neuroscince 17
         (2004), 203-223.
     \bibitem{b:Traub}
         \textit{ R.D. Traub, J.G.R. Jefferys, M.A. Whittington,}
         Fast Oscillations in Cortical Circuits. The MIT Press, Massachusettes,
         1999.
     \bibitem{b:Llinas}
         \textit{Llinas R. and Yarom, Y., } Oscillatory properties
         of guinea-pig inferioir olivary neurines and their
         pharmacological modulation: An in vitro study. J. Physol.,
         Lond., 315, 569-84, 1986.
     \bibitem{b:LlinasVortex}
         \textit{Llinas R., }
         I of vortex. From Neurones  to Self.The MIT Press,
         Massachusettes, 2002.
     \bibitem{b:CazellisCourbage}
         \textit{B. Cazelles, M. Courbage, M. Rabinovich,}
         Anti-phase regularization of coupled
         chaotic maps modelling bursting neurons. Europhysics
         Letters, 56 (4), pp. 504-509 (2001).
     \bibitem{b:RulkovSynchro}
         \textit{N.F. Rulkov, }
         Regularization of synchronized chaotic bursts, Phys. Rev.
         Lett. 86, 183-186 (2001)
     \bibitem{b:RulkovBase}
         \textit{N.F. Rulkov.}
         Modeling of spiking-bursting neural behavior using two-dimensional
         map. Phys. Rev.E, v.65, p. 0.41922 (2002).
     \bibitem{b:RulkovChaos1}
         \textit{A.L. Shilnikov, N.F. Rulkov.}
         Origin of chaos in a two dimensional map modeling
         spiking-bursting neural activity. Int. J. Bifurc. Chaos,
         v.13, N11, pp. 3325-3340 (2003).
     \bibitem{b:RulkovChaos2}
         \textit{A.L. Shilnikov, N.F. Rulkov.}
         Subthreshold oscillations in a map-based neron model. Physics
         Letters A 328, pp. 177-184 (2004).
\bibitem{b:TanakaH} \textit{H. Tanaka}
         Design of bursting in two-dimensional discrete-time neron
         model.Physics Letters A 350, pp. 228-231 (2006).
\bibitem{b:CourbageNekorkin}
         \textit{M.Courbage, V.B. Kazantsev, V.I. Nekorkin, V. Senneret.}
         Emergence of chaotic attractor and anti-synchronization for two coupled monostable neurons. Chaos 12, pp. 1148-1156 (2004).

     \bibitem{b:Izhikevich}
         \textit{E.M. Izhikevich, F. Hoppensteadt, }
         Classification of bursting mappings. Int. J. Bifurcation
         and Chaos, v.14, N11, pp 3847-3854, (2004).
     \bibitem{b:Afraimovich} \textit{V.S. Afraimovich, Sze-Bi Hsu.}  Lectures on
         Chaotic Dynamical Systems, American Mathematical Society.
         Int. Press, 354 p. (2003)
     \bibitem{b:AfraimovichShilnikov} \textit{V.S. Afraimovich, L.P.
         Shilnikov.} Strange attractors and quasiattractors. In
         book "Nonlinear Dynamics and Turbulence" (eds. G.I.
         Barenblatt, G. Iooss, D.D. Joseph, Pitam, Boston, 1983,
         pp. 1-34
     \bibitem{b:Arnold}
         \textit{V.I. Arnold, V.S. Afraimovich, Yu.
             S. Ilyashenko,L.P. Shilnikov.}
         Bifurcation Theory,Dyn. Sys. V. Encyclopaedia Mathematics
         Sciences, Springer, Berlin, 1994
     \bibitem{b:Bernardo}
         \textit{Bernardo L.S., Foster R.P. }
         Oscillatory behaviour in the inferior olive neurons :
         mechanism, modulation, cell agregates. Brain Res. Bull.
         1986, v.17, p.773
     \bibitem{b:Wang}
         R.S.K. Wang and D.A. Prince
         Afterpotential generation in hippocampal pyramidal cells.
         J. Neurophysiol. v45, 1981, p.86
     \bibitem{b:Deschenes}
         M. Deschenes, J.P. Roy and M. Steriade
         Thalamic bursting mechanism: an inward slow current revealed by
         membrane hyperpolarization. Brain Res. 239, 1982, p.289.
\end{thebibliography}
\end{document}